\documentclass[twocolappendix]{emulateapj}
\usepackage{subfigure}
\usepackage{amsmath}

\begin{document}

\slugcomment{Accepted for publication in ApJ: March 31, 2017 (submitted Nov 2016)}

\title{Radial Surface Density Profiles of Gas and Dust in the Debris Disk around 49 Ceti}

\author{A. Meredith Hughes\altaffilmark{1}, 
Jesse Lieman-Sifry\altaffilmark{1},
Kevin M. Flaherty\altaffilmark{1},
Cail M. Daley\altaffilmark{1},
Aki Roberge\altaffilmark{2},
\'Agnes K\'osp\'al\altaffilmark{3},
Attila Mo\'or\altaffilmark{3},
Inga Kamp\altaffilmark{4},
David J. Wilner\altaffilmark{5},
Sean M. Andrews\altaffilmark{5},
Joel H. Kastner\altaffilmark{6},
Peter \'Abrah\'am\altaffilmark{3}
}

\email{amhughes@astro.wesleyan.edu}

\begin{abstract}
We present $\sim$0\farcs4 resolution images of CO(3-2) and associated continuum emission from the gas-bearing debris disk around the nearby A star 49 Ceti, observed with the Atacama Large Millimeter/Submillimeter Array (ALMA).  We analyze the ALMA visibilities in tandem with the broad-band spectral energy distribution to measure the radial surface density profiles of dust and gas emission from the system.  The dust surface density decreases with radius between $\sim100$ and $310$\,au, with a marginally significant enhancement of surface density at a radius of $\sim110$\,au.  The SED requires an inner disk of small grains in addition to the outer disk of larger grains resolved by ALMA.  The gas disk exhibits a surface density profile that increases with radius, contrary to most previous spatially resolved observations of circumstellar gas disks.  While $\sim$80\% of the CO flux is well described by an axisymmetric power-law disk in Keplerian rotation about the central star, residuals at $\sim$20\% of the peak flux exhibit a departure from axisymmetry suggestive of spiral arms or a warp in the gas disk.  The radial extent of the gas disk ($\sim220$\,au) is smaller than that of the dust disk ($\sim300$\,au), consistent with recent observations of other gas-bearing debris disks.  While there are so far only three broad debris disks with well characterized { radial} dust { profiles} at millimeter wavelengths, 49 Ceti's disk shows a markedly different structure from { two radially resolved} gas-poor debris disks, implying that the physical processes generating and sculpting the gas and dust are fundamentally different. 
\end{abstract}

\altaffiltext{1}{Department of Astronomy, Van Vleck Observatory, Wesleyan University, 96 Foss Hill Dr., Middletown, CT 06459, USA}
\altaffiltext{2}{Exoplanets \& Stellar Astrophysics Laboratory, NASA Goddard Space Flight Center, Code 667, Greenbelt, MD 20771, USA}
\altaffiltext{3}{Konkoly Observatory, Research Centre for Astronomy and Earth Sciences, Hungarian Academy of Sciences, P. O. Box 67, 1525 Budapest, Hungary}
\altaffiltext{4}{Kapteyn Astronomical Institute, University of Groningen, Postbus 800, 9700 AV Groningen, The Netherlands}
\altaffiltext{5}{Harvard-Smithsonian Center for Astrophysics, 60 Garden St.,
MS-51, Cambridge, MA 02138, USA}
\altaffiltext{6}{Rochester Institute of Technology, 54 Lomb Memorial Drive, Rochester, NY 14623, USA}

\section{Introduction} 

Planets form in disks of gas and dust around young stars.  Gas-rich ``protoplanetary" disks around pre-main sequence stars are a commonly observed outcome of the star formation process in young star-forming regions, and tend to exhibit inferred masses from millimeter continuum emission that are consistent with the minimum mass that was likely needed to form our Solar System \citep[e.g.][]{and05, and07, ise09, man10, man14, ric14}.  As circumstellar disks evolve their dust masses decrease rapidly, and by ages of $\sim$10-20\,Myr the overwhelming majority of disks have dissipated their primordial dust \citep[e.g.][]{hai01, mam09,bel15}.  Gas dissipation is more difficult to study, but there is some evidence that it proceeds on similar timescales \citep[e.g.,][and references therein]{fed10,den13,wya15}.  The physical mechanisms underlying this dissipation are still an area of active study, but it is likely that the mass is dissipated through a combination of processes including accretion onto the central star, incorporation into planetary systems, and photoevaporative winds launched from the surface of the disk \citep[][and references therein]{ale14}.  At $\gtrsim$10\,Myr ages, the overwhelming majority of remaining circumstellar disks are classified as gas-poor ``debris" disks, in which the small amount of optically thin dust is generated by collisional evolution of Pluto-like bodies in analogs of the Solar System's Kuiper belt \citep[e.g.,][]{wya08}.  Debris disks are a frequently observed phenomenon, detected within current sensitivity limits around a quarter of nearby A stars and 20\% of nearby FGK stars, indicating that planet formation at least to the stage of dwarf planets is commonplace \citep[e.g.][]{eir13,thu14}.  This is supported by observations that planets are a common phenomenon in our galaxy, with an average of at least one planet per star \citep[e.g.,][]{cas12}

While it has been generally assumed that debris disks are devoid of molecular gas, detections of gas emission through millimeter-wavelength spectroscopy date back at least two decades \citep{zuc95,den05}, and absorption line spectroscopy in edge-on debris disks has also identified the presence of a circumstellar gas component around a number of different stars \citep{craw94,rob02,che03,red07}.  Millimeter-wavelength interferometry provided the first spatially resolved observations of the molecular gas emission in a debris disk \citep{hug08}, and the improved sensitivity and angular resolution of the Atacama Large Millimeter/submillimeter Array (ALMA) has led to a resurgence of interest in these objects, resulting in spectacular gas and dust imaging of the recently discovered HD~21997 disk \citep{kos13,moo13}, the iconic $\beta$ Pictoris disk \citep{den14}, and recently the first { gas-bearing} debris disk around a Solar-type star \citep[HD 181327;][]{mar16}.  Another object of interest has been the optically thin dust disk around the 5\,Myr-old Herbig Ae star HD 141569, whose classification as a transitional or debris disk has been an object of intense discussion in the literature: recent observations with the Submillimeter Array (SMA) and Combined Array for Research in Millimeter-wave Astronomy (CARMA) have spatially resolved the gas disk for the first time \citep{fla16}, and ALMA observations probe the disk in greater detail \citep{whi16}.  A survey for gas emission in a sample of 24 debris disks in the 10\,Myr-old Sco-Cen star forming region has revealed that gas-rich debris disks are common around intermediate mass stars, with gas detected around three of the seven A/B stars in the sample, but rare around FGK stars \citep{lie16}.  

The origin of the gas in these debris disks is still a topic of active discussion.  \citet{hug08} originally posited that observations of the 49~Ceti gas disk were consistent with the late stages of photoevaporative clearing of primordial material, but this explanation became less likely once the age estimate for the system was revised from 12\,Myr to 40\,Myr by \citet{zuc12}.  In their study of HD~21997, \citet{kos13} note that while the CO photodissociation timescale in the disk is very short -- of order 10,000\,yr if the disk is H$_2$ depleted, indicating that a primordial origin is unlikely.  { They argue that} the rate of comet destruction needed to sustain the observed gas mass is also unreasonably high, requiring the destruction of 6000 Hale-Bopp-size comets per year (roughly one an hour), { although \citet{zuc12} argue that this rate is not unreasonable if the large observed dust mass of the 49 Ceti system indicates a more substantial reservoir of cometary material than is typical}.  \citet{kos13} also consider shielding of primordial CO by unseen H$_2$ gas in the disk, which could increase the lifetime by up to two orders of magnitude, but this scenario is not obviously consistent with the low excitation temperature of the gas, which appears to indicate departure from LTE and therefore a lack of collision partners in the gas distribution.  

Of the three ALMA-imaged gas disks, $\beta$~Pic stands out as the only one with an obvious departure from axisymmetry, with the southwest limb of the edge-on disk displaying a markedly brighter CO flux than the northeast limb.  \citet{den14} suggest that this asymmetry requires a second-generation origin for the gas, either through a recent collision of Mars-size bodies, or through an enhanced collision rate between grains and small planetesimals concentrated in resonant clumps on either side of the orbit of an unseen $>$10\,M$_\earth$ planet undergoing outward migration through the disk.  The second-generation origin is supported by the model presented in \citet{kra16} an excitation analysis by \citet{mat16} { which favors resonant migration over the giant impact scenario}.

\begin{table*}
\caption{Imaging Parameters for the ALMA observations}
%\resizebox{\textwidth}{!}{
\centering
\begin{tabular}{lccc}
\hline
Wavelength/Line & Beam Size ($''$) & Beam P.A. ($^\circ$) & RMS Noise { (mJy\,beam$^{-1}$)} \\
\hline
850\,$\mu$m Continuum & 0.47$\times$0.39 & 89 & 0.056 \\
HCN(4-3) & 0.54$\times$0.43 & 86 & 3.0 \\
CO(3-2) & 0.50$\times$0.38 & 84 & 8.9 \\
\hline
\end{tabular}
%}
\label{tab:observations}
\end{table*}

\begin{table*}
\caption{Observational parameters for CARMA data}
\centering
\begin{tabular}{cccccc}
\hline
Date & Observation Time & Number of & $\tau_\mathrm{230\,GHz}$ & LO Freq (GHz) & Derived Flux \\
 & on 49 Ceti (hr) & Antennas &  & & of J0132-169 (Jy) \\
 \hline
 2012 Nov 4 & 2.48 & 13 & 0.27-0.37 & 227.5215 & 0.858 \\
 2012 Nov 14 & 1.79 & 13 & 0.17-0.23 & 227.5185 & 0.619 \\
 2012 Nov 15 & 2.23 & 9 & 0.20-0.23 & 227.5184 & 0.659 \\
 2012 Nov 23 & 2.51 & 15 & 0.20-0.23 & 227.5158 & 0.637 \\
 2012 Nov 25 & 2.52 & 14 & 0.17-0.21 & 227.5153 & 0.583 \\
 2012 Nov 27 & 1.56 & 13 & 0.17-0.18 & 227.5147 & 0.600 \\
 2013 Feb 16 & 1.86 & 13 & 0.11-0.26 & 227.5150 & 0.375 \\
 \hline
\end{tabular}
\label{tab:obs_carma}
\end{table*}

With higher angular and spectral resolution observations of the gas disk around 49~Ceti, we now seek clues to the origin of the gas in this system. 49 Ceti is one of only three A stars in the Bright Star Catalogue with a fractional excess luminosity of $L_\mathrm{IR}/L_\mathrm{bol} > 0.001$; the other two are the well-known debris disks HR 4796A and $\beta$ Pic.  However, 49 Ceti has long defied simple classification as a debris disk due to the presence of substantial circumstellar CO emission \citep{zuc95,den05}.  It is remarkable as perhaps the oldest { gas-bearing} disk \citep[40\,Myr;][]{zuc12} and among the closest of the { gas-bearing} disks \citep[{\it Hipparcos} distance 59\,pc;][]{lee07}.  Until recently, very little was known about the distribution of gas and dust in the 49 Ceti system.  {\it HST} coronographic observations { reprocessed with modern imaging techniques  reveal a low surface brightness disk extending from 65 to 250\,au, and imply that the grain size in the outer disk is $>2$\,$\mu$m  \citep[][]{wei99,cho17}}.  Subsarcecond-scale imaging at wavelengths of 12 and 18\,$\mu$m with Keck \citep{wah07} resolved an elongated distribution of small dust grains near the star, but analysis of the SED implied that most of the dust mass in the system lies at larger radii and was not spatially resolved by their observations.  SMA observations spatially resolved the gas disk for the first time in the CO(2-1) line \citep{hug08}, revealing a surprisingly complex and extended molecular gas distribution rotating around the central star, apparently viewed close to edge-on.  The lack of compact, high-velocity emission suggested that the central regions were depleted of CO molecules out to $\sim$90\,au radius (1\farcs5).  In addition, the velocity channels tracing the outer disk exhibited a possible misalignment from the axis defined by the mid-IR observation and the CO tracing the inner disk, perhaps indicative of a warp or a more complex geometry.  Recent {\it Herschel} observations hint at an anomalous elemental abundance.  The 63\,$\mu$m [OI] emission line is undetected, despite the fact that the gas disk model used to fit the SMA CO observations \citep{hug08} would predict a 3$\sigma$ detection.  On the other hand, the {\it Herschel} spectra show a 5$\sigma$ detection of 158\,$\mu$m [CII] emission at a level about 5 times greater than predicted \citep{rob14}.  Thermo-chemical modeling suggests that an anomalous elemental abundance in the gas -- possibly an enhanced C/O ratio -- may be needed to explain these observations.  This may also be the case for the $\beta$ Pic disk gas, which comes from second-generation asteroidal or cometary material \citep{rob06, xie13}, although at least one model is able to reproduce the emission without an anomalous C/O ratio \citep{kra16}.  

Here we present new 0\farcs4 angular resolution observations of gas and dust in the 49~Ceti debris disk observed with ALMA.  We spatially resolve the radial surface density profile of the outer dust disk -- the reservoir of the overwhelming majority of the dust mass -- and simultaneously improve upon previous imaging of the CO disk by a factor of $\sim$3 in spatial resolution and more than an order of magnitude in spectral resolution.  We also present CARMA observations that detect the dust disk at a wavelength of 1.3\,mm, providing reliable constraints on the millimeter spectral slope.  Section~\ref{sec:obs} describes the observations, followed by a discussion of the basic results in Section~\ref{sec:results}.  In Section~\ref{sec:analysis} we describe our efforts to fit both the resolved ALMA  continuum visibilities and the Spectral Energy Distribution (SED) with a self-consistent dust surface density profile, as well as an initial fit to the gas density and temperature profile, based on a simultaneous fit to the ALMA CO(3-2) data and the previously published SMA CO(2-1) observations.  We discuss the implications of our analysis for the origin of gas and dust in the 49~Ceti disk in Section~\ref{sec:discussion}, and summarize our study in Section~\ref{sec:conclusions}.  

\section{Observations} 
\label{sec:obs}

\subsection{ALMA Observations}

49 Ceti was observed with ALMA on 2013 November 4.  The observation incorporated 28 12\,m antennas with baseline lengths between 17 and 1284\,m using the Band 7 receivers under excellent weather conditions ($\sim 0.4$\,mm of precipitable water vapor).  

Observations cycled every 9 minutes between 49 Ceti (40 minutes total on source) and the nearby quasar J0132-1654, which was used as the primary gain calibrator.  Bandpass calibrations were conducted at the beginning of the scheduling block using the quasar J0006-0623, while the absolute flux scale was calibrated using observations of the quasar J0334-401.

The correlator was configured with two broad, low-resolution continuum bands centered at 357 and 344\,GHz, each with the maximum bandwidth of 2\,GHz and channel width 15.6\,MHz to maximize continuum sensitivity.  Another broad band with 1.9\,GHz bandwidth and 488\,kHz channel spacing was centered at 355\,GHz in order to observe the HCN(4-3) transition at a frequency of 354.50547\,GHz; the rest of the band was incorporated into the continuum measurement to improve sensitivity.  The fourth spectral window, with a bandwidth of 234\,MHz and channel spacing of 61\,kHz, was centered on the CO(3-2) transition at a frequency of 345.79599\,GHz.  

The visibilities were calibrated and imaged using the {\tt CASA} and {\tt MIRIAD} software packages, including the standard reduction scripts provided by ALMA.  Initial phase correction was accomplished using the water vapor radiometry system, followed by a system temperature calibration to account for changing atmospheric and instrumental conditions.  After standard bandpass, flux, and gain calibrations, the resulting calibrated visibilities from the three broad-band spectral windows were spectrally averaged to create a single continuum channel with an average frequency of 352\,GHz, corresponding to a wavelength of 850\,$\mu$m.  The continuum, HCN(4-3), and CO(3-2) visibilities were imaged using standard Fourier inversion (assuming natural weighting) and the {\tt CLEAN} algorithm.  Synthesized beam sizes (major and minor axis lengths of the clean beam), position angles, and rms noise levels in the resulting maps { (imaged at native spectral resolution for the two lines)} are listed in Table~\ref{tab:observations}.  The flux scale is subject to a standard 10\% systematic calibration uncertainty.

\subsection{CARMA Observations}

\begin{figure*}[ht!]
\centering
\includegraphics[scale=0.9]{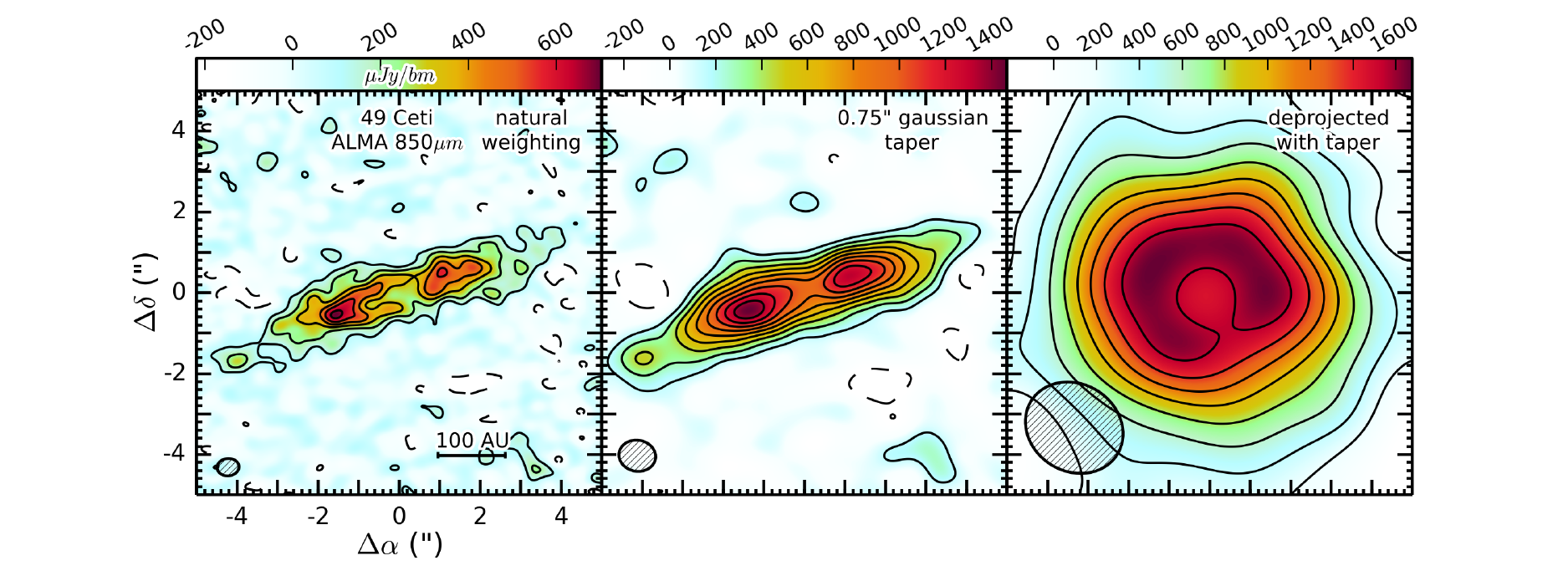}
\caption{
850\,$\mu$m dust continuum emission from the disk around 49 Ceti.  The left panel shows a naturally weighted image of the continuum visibilities.  The center panel shows the same data imaged with a 0\farcs75 Gaussian taper to bring out extended structure in the disk.  The right panel shows a ``deprojected" continuum image \citep[see, e.g.,][]{lay97,hug07}, in which the visibilities have been adjusted to simulate a face-on view of a circular disk with inclination 79.3$^\circ$ and position angle -71.3$^\circ$, consistent with the best-fit geometry of the 49 Ceti disk determined in Section~\ref{sec:analysis_cont}.  In all panels, contours are [-2,2,4,6,...]$\sigma$, where $\sigma$ is the rms noise in the image.  The rms noise for the naturally weighted image is 56{ \,$\mu$Jy\,beam$^{-1}$}, while for the image with the Gaussian taper it is 75{ \,$\mu$Jy\,beam$^{-1}$}, and for the deprojected and tapered image it is 97{ \,$\mu$Jy\,beam$^{-1}$}.  The hatched ellipse in the lower left of each panel represents the synthesized beam, and a scale bar is shown in the bottom right corner of the left panel, assuming a distance of 59\,pc to the 49 Ceti system.  
}
\label{fig:cont_fig}
\end{figure*}

We observed 49 Ceti in the 1.3\,mm continuum with the Combined Array for Research in Millimeter-wave Astronomy (CARMA) over 7 nights in late 2012 and early 2013 for a total of 15 hours.  Our observations used the 15-element subarray that consists of six 10.4\,m diameter antennas and nine 6.1\,m diameter antennas in the compact (D) configuration, providing baseline lengths between 11 and 147\,m.  Table~\ref{tab:obs_carma} contains the date of observation, length of observation per night on target, number of antennas used, approximate opacity at 230\,GHz ($\tau_\mathrm{230\,GHz}$), local oscillator (LO) frequency, and derived flux of the gain calibrator, the quasar J0132-169. 

The weather was adequate, with 230\,GHz opacity varying between 0.11 and 0.37.  Data from malfunctioning antennas and from periods of time when the phase rms varied widely were not used in imaging.  Due to the southern declination of the source relative to the array's location, the synthesized beam was elongated in the north-south direction.  The correlator was configured to take advantage of the maximum continuum bandwidth available, with the LO set to prevent the CO(2-1) line from appearing in any of the 16 spectral windows.  Each window had a bandwidth of 487.5\,MHz, for an aggregate bandwidth of 7.8\,GHz per polarization.  

Each night of observation began with short exposures of Uranus (5\,minutes) to set the flux scale, followed by the bright radio source 3C84 (10\,minutes) to act as a backup flux calibrator and as the bandpass calibrator.  Observations cycled through the quasar J0132-169, 49 Ceti, and J0204-170.  The quasar J0132-169, located 1.3$^\circ$ from 49 Ceti, was used as the gain calibrator.  J0204-170, another nearby quasar, allowed us to assess the quality of the phase transfer from J0132-169.  { A point-source fit to the J0204-170 visibilities using the MIRIAD task {\tt uvfit} returns position offsets of 0\farcs056 in $\alpha$ and 0\farcs15 in $\delta$, indicating that the quality of the phase transfer is good and any absolute astrometric uncertainties are far smaller than the synthesized beam (3\farcs0$\times$2\farcs2 at a position angle of 28$^\circ$).}  A conservative estimate for the systematic uncertainty in the flux scale, set by Uranus, is 10\%.

\section{Results}
\label{sec:results}

\subsection{ALMA Dust Continuum}

Strong continuum emission from 49 Ceti is detected at a wavelength of 850\,$\mu$m.  The left panel of Figure~\ref{fig:cont_fig} shows a naturally weighted image of the continuum emission from the disk.  In order to derive an initial estimate of the integrated flux density, we use the MIRIAD task {\tt uvfit} to fit an elliptical Gaussian to the visibilities on baselines $\le 50$\,k$\lambda$, as these correspond to the largest angular scales in the data and are less affected by the small-scale structure introduced by the gap in the center of the disk.  This method results in a measured total flux density of $17\pm3$\,mJy.  Taking into account the $\sim$10\% systematic flux uncertainty that applies to both data sets, there is no statistically significant difference between the ALMA 850\,$\mu$m flux density and the JCMT/SCUBA2 850\,$\mu$m flux density of 12.1$\pm$2.0\,mJy (\citealt{gre16}; see also \citealt{son04}).  

In order to estimate the geometry of the ring, we also use {\tt uvfit} to fit a thin ring to the visibilities (including all baselines, unlike the flux estimate above).  This fit results in a major axis diameter of 3\farcs822$\pm$0\farcs003 ($233.0 \pm 0.2$\,au) and a minor axis diameter of 0\farcs626$\pm$0\farcs003 ($38.2 \pm 0.2$\,au) at a position angle (defined East of North) of $-70.9^\circ \pm 0.4^\circ$.  If we assume a circular geometry for the disk, this axis ratio implies an inclination to the line of sight of $80.6^\circ \pm 0.4^\circ$, and a radius of approximately 117\,au for the brightest part of the disk.  These values are marginally consistent with those derived from subarcsecond Keck images of thermal emission from the inner disk by \citet{wah07}, who measured a position angle of $-55^\circ \pm 10^\circ$ and inclination $60^\circ\pm15^\circ$.  They are also consistent with the values derived from CO emission in the outer gas disk observed with the Submillimeter Array by \citet{hug08}, who derive a position angle of $-70^\circ \pm 10^\circ$ and inclination $90^\circ \pm 5^\circ$, and with the geometry derived from {\it Herschel} PACS images by \citet{moo15a}, yielding a position angle of -70.6$^\circ \pm$1.2$^\circ$ and inclination 81.4$^\circ \pm$3.9$^\circ$.

\subsection{CARMA Dust Continuum}

\begin{figure}[t!]
\centering
\includegraphics[scale=0.5]{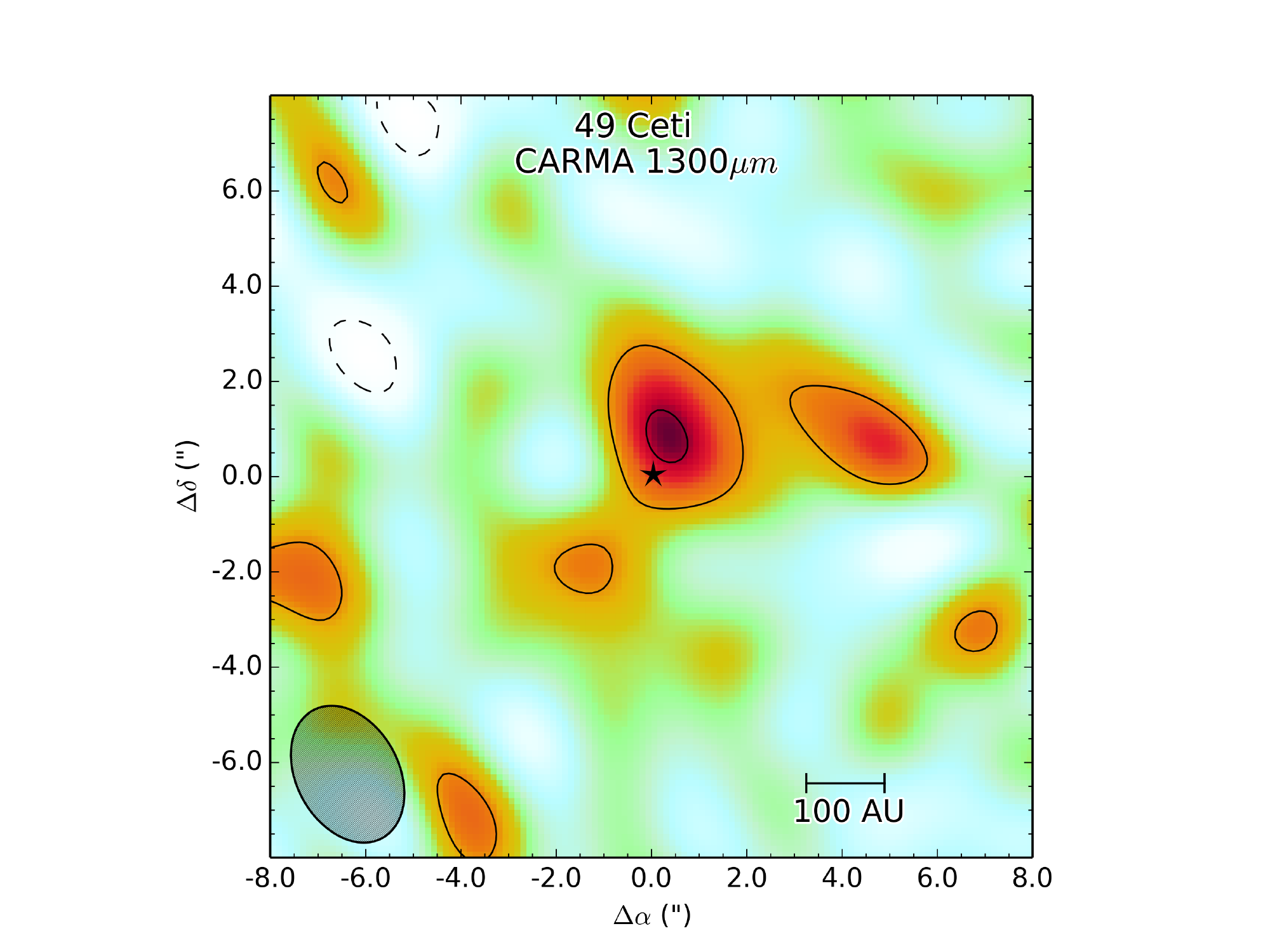}
\caption{
1300\,$\mu$m CARMA image of the dust disk of 49~Ceti.  The contours are [-2, 2, 4] $\times$ 520\,$\mu$Jy\,beam$^{-1}$ (the rms noise).  The synthesized beam size, represented by the shaded ellipse in the bottom left corner, is 2\farcs2$\times$3\farcs0 at a position angle of 28$^\circ$.  
}
\label{fig:carma_49cet}
\end{figure}

Figure \ref{fig:carma_49cet} displays the CARMA image of 49 Ceti at a wavelength of 1300\,$\mu$m.  The combination of seven nights ($\sim$15\,hours) of data resulted in a 4$\sigma$ detection (in the image domain) of flux from the disk at the expected position of the star.  { The phase center of the observations on each night (the origin of the axes in Fig. 2) corresponds to the expected position of the star after accounting for the proper motion of $94.8\pm0.3$\,mas\,yr$^{-1}$ in RA and
$-3.14\pm0.19$\,mas\,yr$^{-1}$ in declination \citep{lee07}.}  The data provide a spatial resolution of $\sim$2\farcs2$\times$3\farcs0 when imaged with a Briggs robust parameter of 0.5.  The disk is apparently unresolved along both the major and minor axis, according to the results of an elliptical Gaussian fit performed with the MIRIAD task {\tt uvmodel}.   We derive a total flux of $2.1 \pm 0.7$\,mJy by fitting a point source to the visibilities with the MIRIAD command {\tt uvfit}.  { In comparison with the size of the continuum ring observed in the ALMA data,
it is somewhat surprising that the CARMA data appear unresolved.  Given the very 
low signal-to-noise ratio of the data, it is impossible at this stage to 
determine whether the morphology of the observed CARMA data reflects a true 
difference in morphology between grains of different sizes or whether it is 
simply the result of sensitivity limitations and random noise in low surface 
brightness data.  The surface brightness of the best-fit model that reproduces 
the ALMA visibilities described in Section 4.1 below predicts a surface 
brightness at CARMA wavelengths that is essentially undetectable outside the 
central 1.5 beams at the sensitivity of the data.  It is therefore possible 
that some extended flux is missed by the CARMA continuum data, and the total flux 
measurement may be an underestimate.}
\clearpage

\subsection{ALMA CO(3-2)}
\label{sec:co_obs}

\begin{figure*}[ht!]
\centering
\plottwo{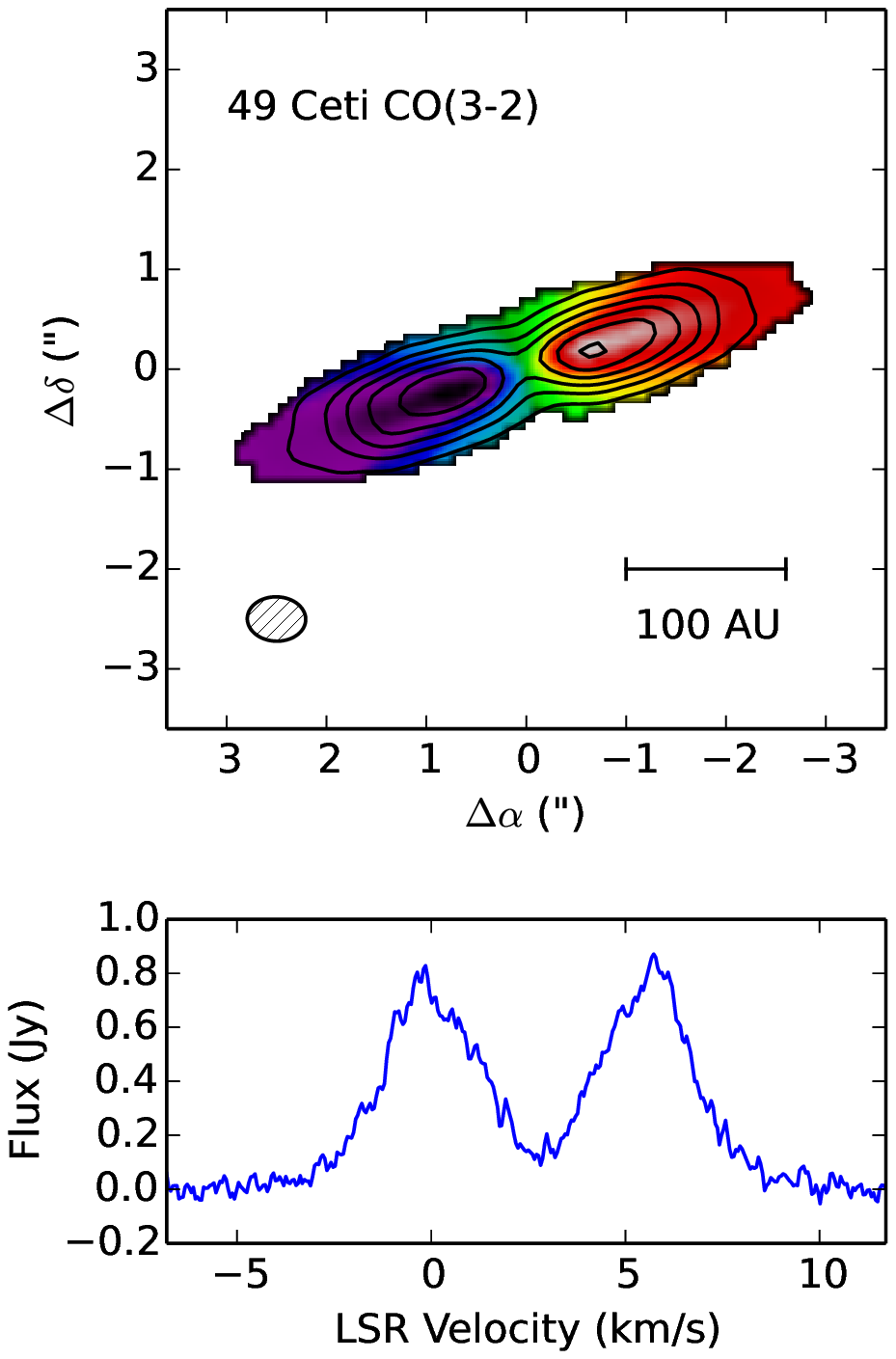}{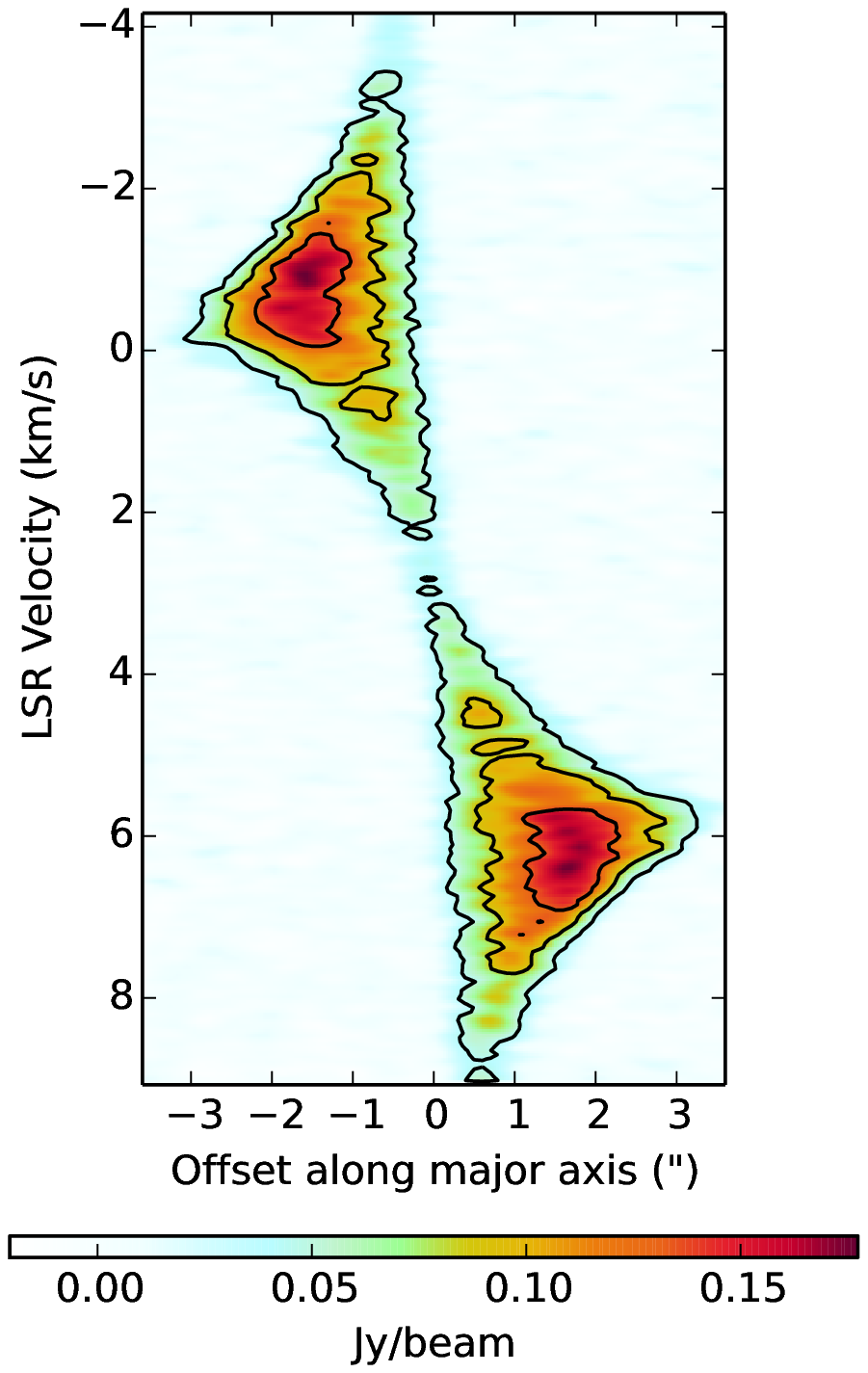}
\caption{
CO(3-2) emission from the disk around 49 Ceti.  The top left panel shows the zeroth (contours) and first (colors) moment maps.  The zeroth moment is the velocity-integrated intensity, while the first moment is the intensity-weighted velocity.  Contours are [5,10,15,...]$\times$0.015\,Jy\,beam$^{-1}$\,km\,s$^{-1}$ (the rms noise in the zeroth moment map).  The hatched ellipse in the lower left represents the synthesized beam. The right panel shows a position-velocity diagram of the CO emission along the disk major axis { (at a position angle of -72$^\circ$)}, starting from the southeast limb of the disk and increasing in position towards the northwest limb.  The color bar illustrates the flux scale in units of Jy\,beam$^{-1}$ { and the contours are [5,10,15,...]$\times$8.9\,mJy\,beam$^{-1}$ (the rms noise in the channel maps)}.  The bottom left panel shows a spatially integrated spectrum of the CO(3-2) emission from the disk at the full spectral resolution of the ALMA data.  The double-peaked zeroth moment map is consistent with a deficit of CO emission in the inner tens of au around the central star { if the emission is optically thin}.  
}
\label{fig:co_fig}
\end{figure*}

We detect strong CO(3-2) emission from the 49 Ceti disk across approximately 250 channels, each with a width of 0.05\,km\,s$^{-1}$.  We illustrate the basic properties of the CO(3-2) emission in Figure~\ref{fig:co_fig}.  The emission is symmetric and regular and appears consistent with Keplerian rotation about the central star.  

We measured the total flux over the region enclosed by the 3$\sigma$ contours in the zeroth moment map using the MIRIAD task {\tt cgcurs}. We derive a total flux density of 6.0$\pm$0.1\,Jy\,km\,s$^{-1}$.  This flux is consistent with the integrated flux from spatially unresolved observations of the CO(3-2) spectrum with the James Clerk Maxwell Telescope (JCMT) by \citet{den05}.  They report an integrated CO intensity of 0.34$\pm$0.07\,K\,km\,s$^{-1}$, which corresponds to an integrated flux of 6.6$\pm$1.4\,Jy\,km\,s$^{-1}$ \citep{mee12}.  Both measurements are subject to the typical $\sim$10\% systematic flux uncertainty at this frequency.  The ALMA observations therefore recover most, if not all, of the CO flux from the disk.  
The maximum width of the 3$\sigma$ contour along the major axis of the disk is 7\farcs3$\pm$0\farcs1, corresponding to an outer diameter of the CO disk of 430\,au at a distance of 59\,pc.  This indicates that the expected outer radius of CO emission from the disk is approximately 220\,au, although the convolution of the beam with the true sky brightness distribution may result in a slight overestimate of this value.  Since most of the flux is recovered, it is unlikely that a significant amount of CO flux extends farther from the disk at a level below the sensitivity of the ALMA map.  Modeling of the visibilities is required to constrain the outer radius more precisely (see Section~\ref{sec:analysis_co} below).  

The peak flux density of the two limbs in the zeroth moment map is identical to within the uncertainties: 0.47$\pm$0.02\,Jy\,beam$^{-1}$ for the northwestern peak and 0.46$\pm$0.02\,Jy\,beam$^{-1}$ for the southeastern peak.  The separation between these peaks is 1\farcs43$\pm$0\farcs03, corresponding to a spatial separation of 84\,au (radius 42\,au) at a distance of 59\,pc.  { If the CO is optically thin,} the lack of a central emission peak in the CO(3-2) moment map is an indication of a central cavity in the CO emission; { however,} for optically thick edge-on disks it is possible to artificially create a double-peaked structure via optical depth effects, as illustrated in \citet{wol08}.  

We perform a simple two-dimensional elliptical Gaussian fit to the visibilities using the MIRIAD task {\tt uvfit} to constrain the position angle of the CO disk.  This method returns a position angle of $-72.0^\circ \pm0.2^\circ$, which is consistent to within the uncertainties with the position angle of the continuum data and previous observations of the CO disk.  It is also consistent with the position angle of $-70.0^\circ \pm 0.5^\circ$ derived from the location of the flux peaks in the CO map along the northwest and southeast limbs of the disk.  { The PV diagram in Fig. 3 represents a single-pixel cut across the disk midplane, at a position angle of -72$^\circ$ and passing through the proper motion-corrected star position at the phase center of the observations.}

\subsection{ALMA HCN(4-3)}

We do not detect any HCN(4-3) emission from the 49 Ceti disk to within the uncertainties in the data.  At the full 0.41\,km\,s$^{-1}$ spectral resolution of the observations, the rms noise in each channel is 3.0\,mJy\,beam$^{-1}$, with a beam size of 0\farcs56$\times$0\farcs43. Since the FWHM of the CO line is 8\,km\,s$^{-1}$, we binned the data by a factor of 5 in the spectral domain to achieve a higher signal-to-noise per channel, and still detected no HCN emission at the $>3\sigma$ level with an rms noise of 1.7\,mJy\,beam$^{-1}$ in each 2.5\,km\,s$^{-1}$-wide channel.  

\section{Analysis} 
\label{sec:analysis}

\subsection{Continuum}
\label{sec:analysis_cont}

We simultaneously model the spatially resolved 850\,$\mu$m ALMA visibilities and the unresolved SED in order to constrain geometric properties of the disk and determine the basic characteristics of the constituent dust grains.  The modeling process and simplifying assumptions are described in detail in \citet{ric13} and \citet{ste16}; we assume a geometrically flat and optically thin dust disk, astrosilicate opacities to determine the temperature of the dust grains as a function of their size, and a simplified emission efficiency of $Q_\lambda = 1 - e^{-(\frac{\lambda}{2\pi a})^{-\beta}}$ \citep[following][]{wil04}.  The hybrid approach of assuming tabulated astrosilicate opacities for the temperature calculation, while also using a parameterized emission efficiency, allows us to approximate the effects of a grain size distribution on the SED while maintaining sufficient computational efficiency to take advantage of the powerful Markov Chain Monte Carlo (MCMC) approach to fitting the data and understanding the uncertainties on the model parameters.  A similar approach is used by \citet{paw14}, who also compare their version of the modified blackbody with a true grain size distribution fit and find that the results are generally consistent.  

The dust grain properties are characterized by two parameters: the characteristic grain size $a$ and long-wavelength power law index of grain emission efficiency $\beta$.  We calculate temperatures on the basis of the characteristic grain size, assuming a compact astrosilicate composition \citep{dra03}.  We determine the opacity, $\kappa_\mathrm{tot}(a,\lambda)$, and albedo, $\omega(a,\lambda)$, of the grains using the Mie theory implementation in MCFOST \citep{pin06,pin09}, and use these properties to calculate the temperature of the dust assuming radiative equilibrium.  

We compare each model to the SED of the 49 Ceti system at observed photometric points gathered from the literature (displayed in Table~\ref{tab:SED}).  We model the SED with two components: (1) a Kurucz-Lejeune model photosphere with surface gravity $\log{g} = 4.5$, effective temperature $T_\mathrm{eff}$ = 10,000\,K, and solar metallicity $Z = 0.01$ \citep{che06}, and (2) an extended, spatially resolved debris disk.  We calculate a $\chi^2$ metric as a test of goodness of fit for the SED using only data points at wavelengths greater than 5\,$\mu$m, since shorter wavelengths are strongly dominated by the stellar photosphere and are not affected by changes to the disk model parameters.  We note that more recent sources derive a lower effective temperature of 8900-9500\,K and slight reddening towards 49~Ceti \citep{mon09,rob13,moo15a}. Since we do not include wavelengths $<$5\,$\mu$m in our fit the details of the model in the optical/UV range do not affect the excess that we derive at longer wavelengths.  The ALMA flux is not included in the calculation of the $\chi^2_\mathrm{SED}$, as this flux is implicitly included in the creation of the model image at 850\,$\mu$m that is compared with the visibilities when calculating the goodness of fit parameter for the visibilities, $\chi^2_\mathrm{vis}$.  The {\it Spitzer} IRS spectrum \citep[taken from the CASSIS database;][]{leb11} originally consisted of 360 points between 5 and 35\,$\mu$m, but was averaged down to 9 for the sake of computational efficiency.  Each of the nine points was included as a single data point (with appropriately weighted errors) when calculating the SED $\chi^2$ value.  The errors reported in the table for these points are the absolute calibration uncertainty (assumed to be { 10\%} of the flux measurement) and the statistical uncertainty added in quadrature.  

In addition to comparing the model to the observed SED, we also create a high-resolution image of the disk at a wavelength of 850\,$\mu$m for comparison with the spatially resolved ALMA data.  The model image is generated with a spatial resolution of approximately 10\% of the spatial scale sampled by the longest baseline in the ALMA data, which corresponds to 3\,au per pixel.  The model images are sampled at the same spatial frequencies as the ALMA data using the MIRIAD command {\tt uvmodel}, allowing us to compare our model disk to the data in the visibility domain where the uncertainties are well characterized.  We then calculate a $\chi^2_\mathrm{vis}$ statistic as a goodness of fit test of the visibilities.  

\begin{table}
\caption{Spatially Unresolved Continuum Fluxes for the 49 Ceti System}
\resizebox{0.5\textwidth}{!}{
\begin{tabular}{ccc}
\hline
\hline
Wavelength ($\mu$m) & Flux (Jy) & Source \\
\hline
0.38 & 8.68 & Yerkes Observatory, \citet{cow69} \\
0.45 & 20.99 & " \\
0.55 & 20.52 & " \\
1.22 & 9.55 & Michigan Curtis Telescope, \\
1.65 & 6.38 & \citet{hou88} \\
2.18 & 3.98 & " \\
3.55 & 1.72 & " \\
4.77 & 0.96 & " \\
5.86 & $0.69\pm0.07$ & IRS Spectrum \\
7.07 & $0.48 \pm 0.05$ & " \\
8.97 & $0.32 \pm 0.03$ & " \\
11.40 & $0.21\pm0.02$ & " \\
13.90 & $0.18 \pm 0.02$ & " \\
17.13 & $0.19\pm0.02$ & " \\
20.90 & $0.21 \pm 0.02$ & " \\
27.21 & $0.35 \pm 0.03$ & " \\
34.00 & $0.6\pm0.1$ & " \\
11.56 & $0.21 \pm 0.02$ & {\it WISE}, \citet{wri10} \\
12.0 & $0.33 \pm 0.07$ & {\it IRAS} Faint Source Catalog \\
25.0 & $0.38 \pm 0.08$ & " \\
60.0 & $2.0 \pm 0.4$ & " \\
100.0 & $1.91 \pm 0.38$ & "\\
12.5 & $0.20 \pm 0.03$ & Keck, \citet{wah07} \\
17.9 & $0.19 \pm 0.03$ & " \\
150.0 & $0.8 \pm 0.5$ & ISO \\
170.0 & $1.1 \pm 0.5$ & " \\
63.19 & $2.01 \pm 0.35$ & {\it Herschel}, \citet{rob13} \\
70.00 & $2.142 \pm 0.058$ & " \\
72.84 & $1.95\pm0.32$ & " \\
78.74 & $1.90 \pm 0.31$ & " \\
90.16 & $1.88 \pm 0.32$ & " \\
145.54 & $1.16 \pm 0.18$ & " \\
157.68 & $0.98 \pm 0.13$ & " \\
160.00 & $1.004 \pm 0.053$ & " \\
250.0 & $0.372 \pm 0.027$ & " \\
350.0 & $0.180\pm 0.014$ & " \\
500.0 & $0.086 \pm 0.009$ & " \\
1300.0 & $0.0021 \pm 0.0007$ & CARMA, This work \\
\hline
\end{tabular}
}
\label{tab:SED}
\end{table}

\subsubsection{Surface Density Profiles}

Our primary goal in fitting the ALMA data is to characterize the surface density structure of the 49 Ceti debris disk, based on the spatially resolved ALMA visibility data.  The most common parameterization of surface density in debris disks is a simple power law function of radius, with $\Sigma(r) \propto r^{-p}$ and abrupt cutoffs at the inner and outer radii.  For this model, we need to specify four variables to generate the surface density profile: the total disk mass, $M_\mathrm{disk}$; the surface density power law index $p$; the inner radius, $R_\mathrm{in}$; and the outer radius, $R_\mathrm{out}$.  If we normalize to a radius of 100\,au, the surface density profile is then
\begin{eqnarray}
\Sigma(r) = \Sigma_\mathrm{100\,au} \left(\frac{r}{\mathrm{100\,au}} \right)^{-p}
\end{eqnarray}
where 
\begin{eqnarray}
\Sigma_\mathrm{100\,au} = \frac{M_\mathrm{disk} (2-p)}{2\pi (\mathrm{100\,au})^p (R_\mathrm{out}^{2-p} - R_\mathrm{in}^{2-p})}. 
\end{eqnarray}

However, the appearance of the naturally weighted image of the 49 Ceti disk in Figure~\ref{fig:cont_fig}, which displays a relatively gradual increase and decrease of surface brightness along the semimajor axis, indicates that perhaps a more complex surface density structure is justified.  Motivated by the three models used in the fit to ALMA observations of the debris disk around HD~107146 by \citet{ric15}, we also attempt to fit the data with a radially unresolved surface density enhancement (i.e., a thin ring of increased surface density located at a radius $R_\mathrm{belt}$), and a double power law model in which the surface density profile changes its power law index $p$ at a transition radius $R_\mathrm{T}$ that is constrained to lie between $R_\mathrm{in}$ and $R_\mathrm{out}$.  The unresolved surface density enhancement requires the addition of two parameters to the default single power law model: the radius of the extra belt, $R_\mathrm{belt}$, and the mass of the belt, $M_\mathrm{belt}$.  We model the width of the belt to be 3\,au across the radial dimension, which was chosen to be a small fraction of the spatial scale probed by the longest baseline in the ALMA data set. For the double power law model, we specify three new parameters: the transition radius, $R_\mathrm{T}$, the power law for the inner surface density profile, $p_1$, and the power law for the outer surface density profile, $p_2$.  The surface density profile is then
\begin{eqnarray}
\Sigma(r) = \begin{cases}
\Sigma_\mathrm{T} \left(\frac{r}{R_\mathrm{T}} \right)^{-p_1} \mathrm{~for~} R_\mathrm{in} \le r \le R_\mathrm{T} \\
\Sigma_\mathrm{T} \left(\frac{r}{R_\mathrm{T}} \right)^{-p_2} \mathrm{~for~} R_\mathrm{T} \le r \le R_\mathrm{out}
\end{cases}
\end{eqnarray}
where 
\begin{eqnarray}
\Sigma_\mathrm{T} = \frac{M_\mathrm{disk} j_1 j_2}{2\pi (m_1 j_2 R_\mathrm{T}^{p_1} + m_2 j_1 R_\mathrm{T}^{p_2})}
\end{eqnarray}
and $j_1 = 2 - p_1$, $j_2 = 2 - p_2$, $m_1 = R_\mathrm{T}^{j_1} - R_\mathrm{in}^{j_1}$, and $m_2 = R_\mathrm{out}^{j_2} - R_\mathrm{T}^{j_2}$.  

The three surface density profiles are illustrated schematically in the right column of Figure~\ref{fig:superplot}, where dark green represents high surface density.  To ensure that $R_\mathrm{in} \le R_\mathrm{T} \le R_\mathrm{out}$, within the code we specify the transition radius $R_\mathrm{T}$ as $R_\mathrm{in} + \Delta R_\mathrm{T}$ and the outer radius $R_\mathrm{out}$ as $R_\mathrm{in} + \Delta R_\mathrm{T} + \Delta R_\mathrm{out}$.  

\subsubsection{MCMC Fitting}

In order to find the best fit to the data and explore the uncertainties (including degeneracies) associated with each parameter, we utilize the affine-invariant MCMC fitting technique as described by \citet{good10} and implemented in Python as {\tt emcee} by \citet{fore13}.  We fit the sum of the SED and visibility $\chi^2$ values, { assuming uniform priors on all parameters (although the assumption of uniform priors on those parameters sampled logarithmical is equivalent to the assumption of a log-normal prior).}  Using an MCMC method allows us to probabilistically sample the full parameter space described by our models, obtain a best-fit result, and place uncertainties on the model parameters by characterizing the full posterior distribution function for each parameter.  Several runs were attempted with different starting positions to ensure that the solution was a global rather than a local minimum.  Most of our MCMC chains included approximately 100 walkers that were allowed to move for 1000 steps, and the ``burn-in" phase during which the $\chi^2$ values rapidly decreased as the walkers approached the best-fit region of parameter space typically comprised of order $\sim100$ steps.  The burn-in phase was discarded from each MCMC run.  

We experimented with fitting the spatially resolved visibilities and SED both simultaneously and independently.  These runs made it clear that it was not possible to fit both the SED and visibilities with only a single characteristic dust grain size, $a$, and long-wavelength emission efficiency power law, $\beta$.  For example, using the Akaike Information Criterion (AIC) test, the best-fit simultaneous SED and visibility model using a single power law surface density distribution was a poorer fit to the visibilities than the best visibility-only fit at the $>10\sigma$ level.  This is unsurprising, since comparisons of the mid-IR imaging \citep{wah07} with the ALMA image in Figure~\ref{fig:cont_fig} reveal two spatially disparate grain distributions: the small grains that dominate the thermal IR emission are concentrated within 30-60\,au from the star, while the large grains that dominate the millimeter-wavelength emission exhibit peak surface brightness at a radius of more like 100\,au from the star and extend out to several hundred au.  These observations are consistent with modeling results described in \citet{wah07}, which indicate the need for two grain populations to reproduce both the SED and the resolved imaging in the mid-IR.  We therefore introduce an ``inner disk" component into our model, with a characteristic grain size of $a = 0.1$\,$\mu$m to maintain consistency with \citet{wah07}, variable inner radius $R_\mathrm{Inner~disk}$, and an outer radius that coincides with the inner radius of the outer disk, $R_\mathrm{In}$.  Since the inner disk surface density profile is not spatially resolved enough to constrain the power law index as a function of radius in the \citet{wah07} data, we assume $\Sigma_\mathrm{Inner~disk} \propto r^{-1}$ and vary the total mass of the inner disk $M_\mathrm{Inner~disk}$ and the total mass of the outer disk $M_\mathrm{Outer~disk}$ separately.  For the model with the unresolved surface density enhancement, we also separately vary the mass of the ring, $M_\mathrm{Ring}$.  We find that $R_\mathrm{in}$ is essentially unconstrained by the spatially resolved ALMA data in the case of the double power law (since the spatial resolution is not high enough to distinguish between a sharp and gradual inner disk edge), so we therefore fix the location of $R_\mathrm{in}$ to be { 20\,au} greater than $R_\mathrm{Inner~disk}$ for the double power law model.  

Table~\ref{tab:cont_best_fit} reports both the median and $\pm 1\sigma$ intervals from the posterior distribution, as well as the global best-fit set of model parameters from each MCMC run, for the three classes of surface density structure that were used for the fits.  The best-fit values are generally within $\pm 1 \sigma$ of the median.  For all models, disk mass parameters ($M_\mathrm{Inner~disk}$, $M_\mathrm{Outer~disk}$, and $M_\mathrm{Ring}$) and the grain size are sampled logarithmically, while the rest of the parameters are sampled linearly.  { It should be noted that since the parameter $\beta$ is constrained primarily by the slope of the SED between the ALMA and CARMA flux points, the low sensitivity of the CARMA data to extended low-surface-brightness flux might bias our measurement of $\beta$ to be steeper than the true value.}

\begin{figure}[t!]
\centering
\includegraphics[scale=0.8]{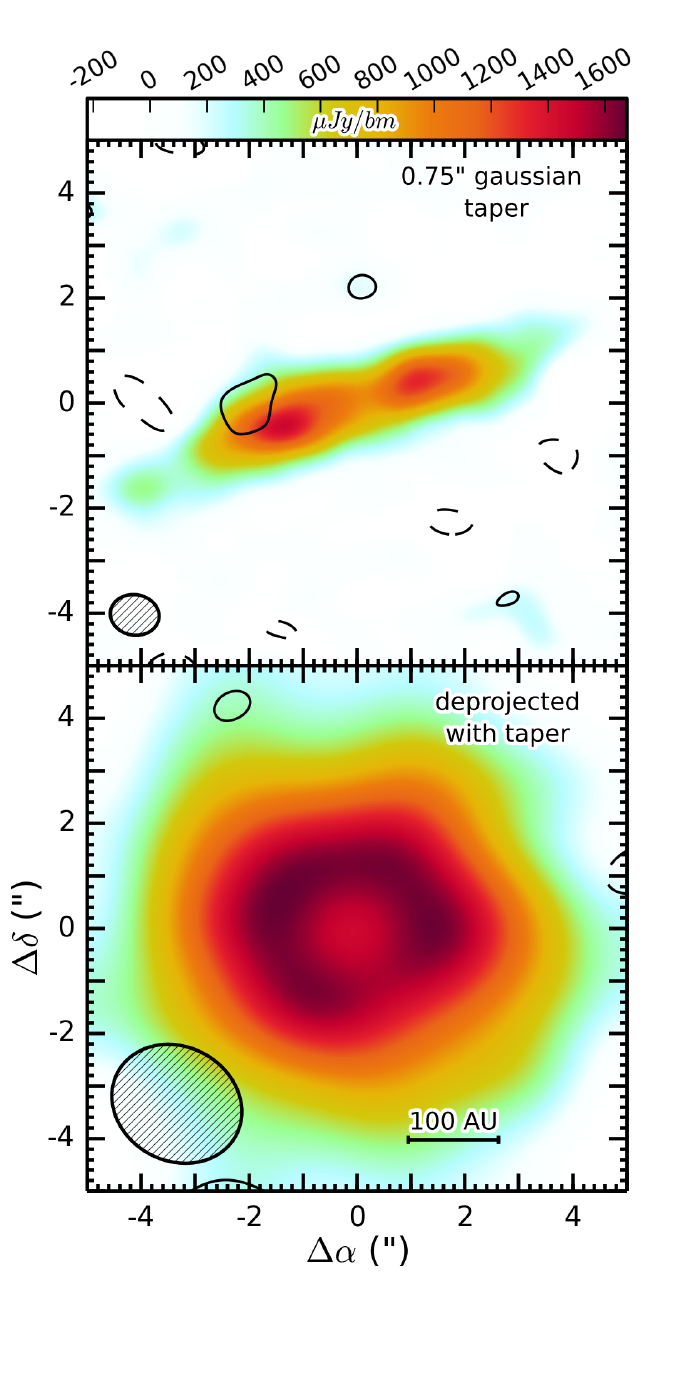}
\caption{Residuals (contours) overlaid on the tapered and deprojected continuum emission from the 49 Ceti disk.  Contour levels are as in Fig.~\ref{fig:cont_fig}.  The lack of statistically significant residual features in the tapered image indicates that there are no statistically significant deviations from axisymmetry even on the scale of a few beams rather than the 0\farcs4 scale of the naturally weighted image.}
\label{fig:taper_resid}
\end{figure}

\begin{figure*}
\centering
\includegraphics[scale=0.6]{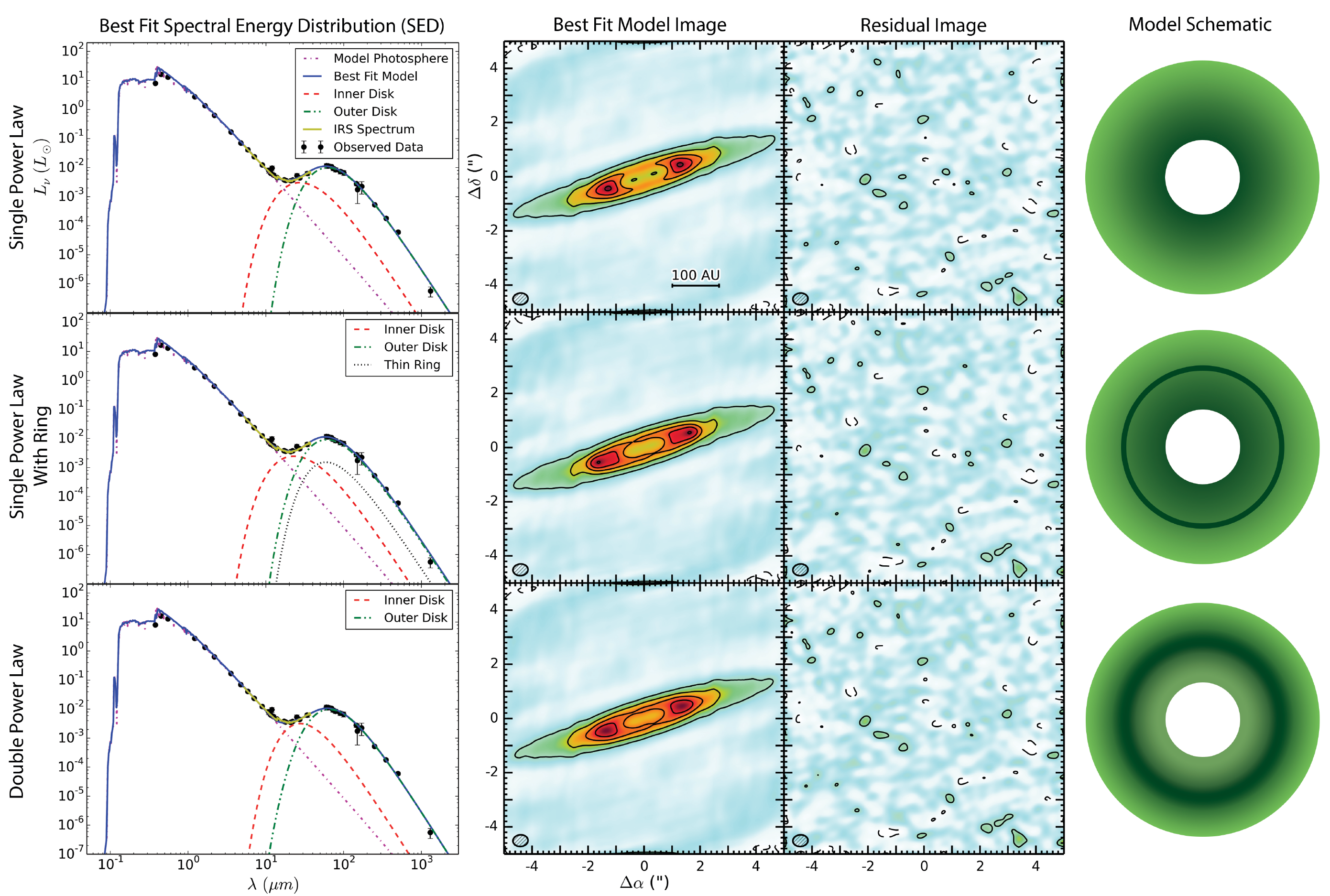}
\caption{
Comparison of data with three different classes of models: The top row shows results for the best-fit single power-law disk model, the center row shows the same comparison for a power-law disk model with an extra spatially unresolved surface density enhancement, and the bottom row shows a model in which the disk is composed of two contiguous power law density distributions that are joined at a transition radius.  The left column shows the observed SED of the 49 Ceti system (black dots) and the {\it Spitzer} IRS spectrum (yellow solid line) compared with the model stellar photosphere (purple dot-dashed line) and the model circumstellar disk (see legends for components of three different models).  The image column in the center shows a simulated ALMA observation of the best-fit model for each of the three types of models (left) and the residual image when the model is subtracted from the data in the visibility domain (right).  In this column, the beam size and shape are illustrated by the hatched ellipse in the bottom left of each panel, and the linear scale of the data is indicated by a bar in the upper left panel. Contour levels { and CLEAN parameters are identical to those} in Fig.~\ref{fig:cont_fig}.  The right column shows a schematic illustration of a face-on view of each of the three classes of model disks (white is highest density, green is lowest density, and the black circle at the center represents the inner disk).  
}
\label{fig:superplot}
\end{figure*}

\begin{figure}[ht!]
\centering
\includegraphics[scale=0.45]{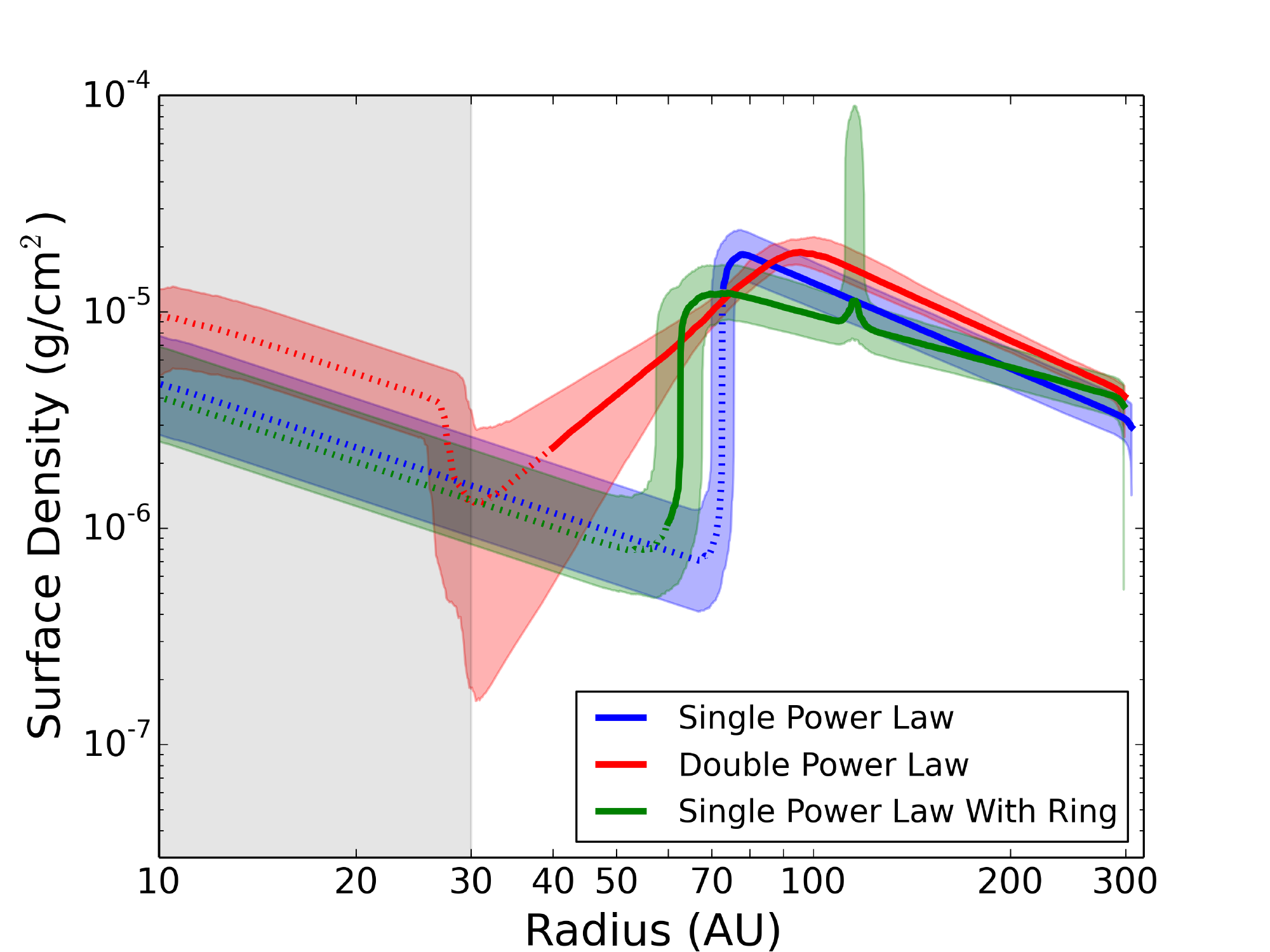}
\caption{
Best-fit surface density profiles for the three classes of models: single power law (blue), double power law (red), and power law with ring (green).  The dotted line indicates the extent of the inner disk \citep[grain size 0.1\,$\mu$m, resolved in Keck imaging by][]{wah07}, while the solid line indicates the extent of the outer disk.  The shaded regions on either side of each line indicate $\pm1\sigma$ uncertainties on the surface density derived from the MCMC posterior distributions.  The gray shaded region to the left of the plot indicates the spatial resolution of the data; spatial scales within the gray shaded region are extrapolated from larger radii.  
}
\label{fig:surf_dens}
\end{figure}

Figure~\ref{fig:superplot} displays a graphical summary of the fitting results, including the best-fit model and residuals for each of the three classes of models.  A schematic of each model is presented in the right column.  The left column compares the observed SED (black points; see Table~\ref{tab:SED}) with the various components of the model SED and the {\it Spitzer} IRS spectrum.  All three of the models are consistent with the data to within the uncertainties, as illustrated by the absence of any statistically significant ($>3\sigma$) features in the residual image.  This lack of residual flux also indicates that the surface density structure of the dust disk is consistent with axisymmetry, despite the slight apparent mismatch of peak flux levels in the ansae of the image derived from the ALMA data (Fig.~\ref{fig:cont_fig}).  { Fig.~\ref{fig:taper_resid} emphasizes this point by plotting the tapered residuals, which also show no significant features, as would be expected if there were a more subtle but significant asymmetry on the scale of a few beams rather than the 0\farcs4 scale of the naturally weighted image.}  Plots of the best-fit surface density profile for the three classes of models are presented in Figure~\ref{fig:surf_dens}.  The three models are consistent in showing an overall trend of decreasing surface density with radius, with the inner radius of the outer disk located between $\sim$50-70\,au and the outer radius close to 300\,au (the inner and outer radius are most uncertain with the double power-law model, due to degeneracy between the slope of the power law and the location of the edges).  The double power law and the single power law with a ring both indicate a maximum in the surface density profile that falls at a radius of $\sim$90-110\,au.  

While the two-dimensional map of the model and residuals indicates no significant asymmetry, subtle differences in the distribution of flux are evident in the three model images, and it is possible that the ensemble of visibilities may be able to distinguish statistically between the three modeled surface density distributions (akin to boosting the signal-to-noise ratio by considering an azimuthally averaged version of the surface density distribution rather than only point-to-point variations in flux).  To explore this possibility, we employ { both the AIC and the related Bayesian Information Criterion (BIC)}, which allow us to make statistical comparisons between the likelihood functions of models with different numbers of parameters.  A key assumption of the AIC is that the posterior distribution functions are Gaussian or nearly Gaussian \citep{lid07}; { the BIC is a similar test that typically imposes a larger penalty for additional free parameters.}  All of the parameters from the MCMC chain are well constrained and fill out  posterior distributions that fit the criteria.  { According to the AIC, there is a marginally significant improvement in the fit to the data when a local surface density enhancement is included in the model: the double power law model is a better fit than the single power law model at the 2.2$\sigma$ level, while the power law with a thin ring is a better fit than the single power law model at the 2.9$\sigma$ level.  By contrast, the AIC detects no significant difference between the double power law and the single power law with a ring (0.5$\sigma$).  The BIC returns similar results, except that $\Delta$BIC is $>10$ when comparing the single power law fit with both the double power law and the power law with a ring, corresponding to ``very strong'' evidence that the models that include an enhancement in density at a radius of $\sim100$\,au.  By contrast, $\Delta$BIC is 4 for the comparison between the double power law and the single power law with a ring, which falls into the category of ``positive'' but not ``strong'' evidence that the power law with a ring is preferred.  There is also a philosophical point to be made that the tendency of both the double power law model and the single power law with a ring to converge on a solution with a density increase at $\sim$100\,au is suggestive of a density enhancement, since either model could instead have converged to the single power law solution.  On the whole, we take these statistical tests to indicate that} models of the surface density that include a surface density enhancement at a distance of $\sim$100\,au from the central star better represent the physical nature of 49 Ceti's dust disk than a single power law.  { The current data cannot distinguish between a radially broad enhancement like the double power law model and a radially narrow enhancement like the power law plus ring.}

\begin{table*}[ht!]
\caption{Best-fit parameters for three models of the dust disk}
\centering
\begin{tabular}{|l|cc|cc|cc|}
\hline
 & \multicolumn{2}{|c|}{Single Power Law} & \multicolumn{2}{c|}{Power Law w/Ring} & \multicolumn{2}{c|}{Double Power Law} \\
Parameter & Median  $\pm 1\sigma$ & Best Fit  & Median  $\pm 1\sigma$ & Best Fit  & Median  $\pm 1\sigma$ & Best Fit  \\
\hline
$R_\mathrm{Inner~Disk}$ [au] & $5_{-1}^{+2}$ & 6 & $4.6_{-1.2}^{+1.5}$ & 3.5 & $7.0_{-1.5}^{+2.0}$ & 6.7 \\
$R_\mathrm{In}$ [au] & $73_{-3}^{+3}$ & 73 & $62_{-5}^{+5}$ & 60 & -- & 26.7 \\
$\Delta R_\mathrm{T}$ [au] & -- & -- & -- & -- & $68^{+7}_{-11}$ & 68.6 \\
$\Delta R_\mathrm{Out}$ [au] & $310^{+9}_{-9}$ & 310 & $302^{+9}_{-8}$ & 303 & $215^{+12}_{-9}$  & 219 \\
$R_\mathrm{Ring}$ [au] & -- & -- & $114^{+3}_{-3}$  & 114 & -- & -- \\
$\log ( M_\mathrm{Inner~disk}$ [M$_\earth$]) & $-3.1^{+0.2}_{-0.2}$ & -3.0 & $-3.3^{+0.2}_{-0.2}$ & -3.4 & $-3.33^{+0.14}_{-0.19}$ & -3.40 \\
$\log ( M_\mathrm{Outer~disk}$ [M$_\earth$]) & $-1.17^{+0.08}_{-0.10}$ & -1.16 & $-1.21^{+0.09}_{-0.10}$ & -1.21 & $-1.07^{+0.04}_{-0.05}$ & -1.09 \\
$\log ( M_\mathrm{Ring}$ [M$_\earth$]) & -- & -- & $-2.26^{+0.15}_{-0.18}$ & -2.26 & -- & -- \\
$\log (a [\mu$m]) & $0.08^{+0.11}_{-0.14}$ & 0.09 & $0.09^{+0.10}_{-0.13}$ & 0.06  & $0.23^{+0.05}_{-0.05}$ & 0.22 \\
$\beta$ & $1.17^{+0.05}_{-0.05}$ & 1.18 & $1.17^{+0.05}_{-0.05}$ & 1.17 & $1.23^{+0.03}_{-0.03}$ & 1.22 \\
$p$ & $1.29^{+0.11}_{-0.11}$ & 1.27 & $0.80^{+0.19}_{-0.17}$ & 0.75 & -- & -- \\
$p_1$ & -- & -- & -- & -- & $-2.5^{+0.8}_{-2.2}$ & -2.7\\
$p_2$ & -- & -- & -- & -- & $1.44^{+0.15}_{-0.20}$ & 1.50\\
$i$ [$^\circ$] & $79.4^{+0.4}_{-0.4}$ & 79.3 & $79.3^{+0.4}_{-0.4}$ & 79.3 & $79.0^{+0.2}_{-0.2}$ & 79.2 \\
$PA$  [$^\circ$] & $-71.4^{+0.4}_{-0.5}$ & -71.3 & $-71.4^{+0.4}_{-0.4}$ & -71.2 & $-71.6^{+0.4}_{-0.5}$ & -71.7 \\
\hline
\end{tabular}
\label{tab:cont_best_fit}
\end{table*}

\subsection{CO(3-2)}
\label{sec:analysis_co}

In order to understand the underlying temperature and density structure of the gas disk, we fit the new ALMA CO(3-2) visibilities in tandem with previously published SMA CO(2-1) observations of the gas disk \citep{hug08}.  The combination of two lines provides constraints on the excitation temperature of the gas in the disk, while the superior sensitivity and angular resolution of the ALMA data provides the best constraints on the brightness profile and kinematics of the gas disk.  

Because previous observations of { gas-bearing} debris disks have derived low excitation temperatures \citep[e.g.,][]{kos13,fla16}, there is a high probability that the disk is not in local thermodynamic equilibrium \citep[LTE; see also][]{mat15}.  { The integrated line ratio of 3.0 in the 49 Ceti system, corresponding to a disk-averaged excitation temperature of 32\,K, is also consistent with a non-LTE scenario.}  We therefore model the gas disk using the non-LTE Line Modeling Engine (LIME) radiative transfer code written by \citet{bri10}.  Since each set of model parameters takes of order an hour to converge, it is not feasible to fit the data using the MCMC method that we employed for the continuum emission.  Instead, we carry out a simple $\chi^2$ minimization via a grid search technique, which is good at locating the best-fit combination of model parameters in a well-behaved space, but does not efficiently characterize the uncertainties on each parameter.  To avoid local minima, we start by searching a broad swath of parameter space using a coarsely sampled grid, and gradually narrow down our search to a smaller, more finely sampled grid.  The $\chi^2$ value that is minimized is the sum of the raw $\chi^2$ values for the SMA and ALMA data.  

{ Since the underlying H$_2$ abundance is unknown, we run two sets of models: the first assumes a constant, nominal CO/H$_2$ abundance of $10^{-4}$, consistent with ISM abundances.  The second assumes a constant CO/H$_2$ abundance of 1, intended to be representative of a secondary gas origin.  We recognize that the chemistry of the 49 Ceti disk could in fact deviate substantially from these values, but a comparison between the model outputs should provide some guidance as to which results are independent of assumptions about the origin of the gas in the system. }

We assume a simple power-law index for the temperature structure of the CO disk, and a similarity solution for the surface density \citep[see, e.g.,][]{hug08b}.  { In cases where the CO emission is optically thin}, the power law indices describing the radial variations of temperature and density may be degenerate.  We therefore assume that the temperature decreases with distance from the star $R$ as $1/\sqrt{R}$, which is a standard assumption based on blackbody equilibrium calculations that may or may not apply to the unusual physical conditions in the disk around 49 Ceti { (we relaxed this assumption for the condition of low H$_2$ abundance, in which the emission is more optically thick, although the best-fit value for the temperature power law remained at 0.5)}.  We also assume that the temperature structure of the gas is vertically isothermal, and depends only on distance from the star: $T(R) = T_{100} \left( \frac{R}{\mathrm{100~au}} \right) ^{-0.5}$. { For the case of CO/H$_2$ = 10$^{-4}$} we use the default assumption of a standard ISM composition for the gas that is 80\% molecular hydrogen by mass with a mean molecular weight $m_0 = 2.37 m_H$, where $m_H$ is the mass of the hydrogen atom; { for the case of CO/H$_2$ = 1 we assume a mean molecular weight of 14, consistent with the dominant mass component being the C and O atoms released from CO photodissociation \citep{mat16}}.  We then calculate the sound speed as a function of radius: $c_s(R) = \sqrt{\frac{k_B T(R)}{m_0}}$.  The scale height is then $h(R) = \frac{\sqrt{2} c_s(R)}{\Omega(R)}$, where $\Omega(R) = \sqrt{\frac{G M_*}{R^3}}$ and $M_*$ is the stellar mass.  We parameterize the surface density $\Sigma$ as a function of radius as 
\begin{eqnarray}
\Sigma(R) = \frac{M_\mathrm{disk}(2-\gamma)}{2\pi R_c^2} \left(\frac{R}{R_c}\right)^{-\gamma} \exp \left[-\left(\frac{R}{R_c}\right)^{2-\gamma}\right], 
\end{eqnarray}
where $R_c$ is the characteristic radius and $\gamma$ describes the power-law drop off of surface density with radius when $R \ll R_c$.  We also include an inner cutoff in the disk surface density at radius $R_\mathrm{in}$.  The volume mass density as a function of radius and height above the midplane $z$ is then 
\begin{eqnarray}
\rho(R,z) = \frac{\Sigma(R)}{\sqrt{\pi} h(R)} \exp{[-\left(\frac{z}{h(R)}\right)^2}].
\end{eqnarray}  
{ We do not include a prescription for CO freeze-out since the freeze-out timescale is inversely proportional to the surface area of grains per unit volume \citep{flo05}, which is likely to be extremely small in the low-density environment of the debris disk around 49 Ceti.}  We further assume a constant turbulent linewidth of 10\,m\,s$^{-1}$, which is substantially smaller than the spectral resolution of the data, and cylindrical Keplerian rotation where $v(R) = \sqrt{\frac{G M_*}{R}}$.  

Initially we varied the following parameters: \{$M_*$, $M_\mathrm{disk}$, $R_c$, $R_\mathrm{in,gas}$, $\gamma$, $T_{100}$, $i$\}.  To minimize the number of free parameters we fixed the position angle $PA$ at the best-fit value of -71.4$^\circ$ from the fits to the continuum visibilities, which is intermediate between the two best-fit CO values of -72.0$^\circ$ (from the two-dimensional Gaussian fit to the integrated CO visibilities) and -70.0$^\circ$ (from the position of the two flux peaks in the CO map).  After the initial coarse $\chi^2$ minimization, we performed a one-dimensional $\chi^2$ minimization on the systemic velocity, considering values between 2 and 3.5\,km\,s$^{-1}$.  We settle on a best-fit LSR velocity of $2.78\pm0.13$\,km\,s$^{-1}$.  

Strong residuals visible along the extreme edges of the disk minor axis suggested that the vertical extent of the CO emission was perhaps incorrect, for both classes of models.  We therefore introduced an additional multiplicative factor $h_0$ that modifies the scale height so that $h(R) = h_0 \frac{\sqrt{2} c_s(R)}{\Omega{R}}$.  The scale height factor $h_0$ is a simple way of accounting parametrically for a disk structure that may be flatter or puffier than expected based on the standard Gaussian scale height approximation to hydrostatic equilibrium.  A physical motivation for this additional factor is that, especially for a disk that is not in LTE, the kinetic gas temperature may differ significantly from the excitation temperature, causing us to calculate an inappropriate scale height since the temperature of the model is constrained by the line ratio rather than the kinematics.  Furthermore, the Gaussian approximation to the vertical structure of the disk is not rigorous, and to be truly internally self consistent we would have to iteratively calculate the vertical structure by solving the equations of hydrostatic equilibrium for each new disk structure that is proposed, which would increase the computational resources required even further.   Introducing this parameter decreased the $\chi^2$ value of the best fit by 2117 to 2598848 (reduced $\chi^2$ 1.35), which represents a significantly better fit at the $>10\sigma$ level according to the AIC test.  

\begin{figure*}
\centering
\includegraphics[scale=0.5]{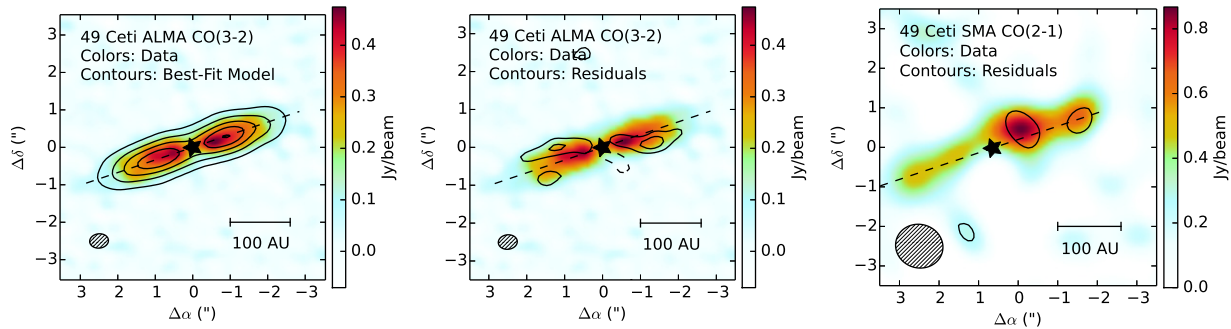}
\caption{
{ Zeroth moment (velocity-integrated intensity) maps of the data, compared with the best-fit model of the ALMA CO(3-2) map (left panels) and SMA CO(2-1) map (right panel).  The colors represent the data in all three panels.  The left panel shows the best-fit model (contours) compared with the data.  Contours represent [5,10,15,...]$\times \sigma$, where $\sigma$ is the rms noise of 0.02\,Jy\,beam$^{-1}$.  The center and right panels show the imaged residuals (contours) after the best-fit model is subtracted from the data in the visibility domain.  }The contours represent [-2,2,4]$\times \sigma$, where $\sigma$ is the RMS noise of 0.02\,Jy\,beam$^{-1}$ (ALMA) and 0.03\,Jy\,beam$^{-1}$ (SMA).  The synthesized beam dimensions are indicated by the hatched ellipse in the lower left corner of each panel.  A dashed line has been added to show the assumed position angle of -71.4$^\circ$, adopted from the continuum and consistent with the location of the flux peaks in the CO(3-2) map as well as a two-dimensional Gaussian fit to the CO visibilities.  The CO(2-1) map is centered on the J2000 coordinates of the star as in \citet{hug08}, so an offset equivalent to the SIMBAD proper motion has been added to the coordinates of the disk major axis.  The residuals in the ALMA map, at roughly 20\% of the peak CO flux, indicate that the CO emission is twisted relative to the disk major axis, potentially indicative of spiral arms, a warp, or other deviation from axisymmetry in the gas disk.  
}
\label{fig:co_resid}
\end{figure*}

In the final fit, we therefore varied the following parameters: \{$M_*$, $M_\mathrm{disk}$, $R_c$, $R_\mathrm{in,gas}$, $\gamma$, $T_{100}$, $i$, $h_0$\}. The best-fit values, minimum and maximum grid values considered, and the finest step size in the final grid are presented in Table~\ref{tab:best_fit_co}.  { The reported best-fit disk masses are the sum of the total H2 
and CO mass.} This best fit results in a reduced $\chi^2$ value of 1.166 for the SMA CO(2-1) data and 1.094 for the ALMA CO(3-2) data { (the differences between the CO/H$_2$ = 10$^{-4}$ and CO/H$_2$ = 1 cases are insignificant at this level of precision)}.  The derived stellar mass of 2.1\,M$_\sun$ is in good agreement with that derived by \citet{rob13} using isochrone fitting.  
Figure~\ref{fig:co_resid} presents the residual moment zero map { for the case of CO/H$_2$ = 10$^{-4}$} (contours) overlaid on the moment zero map of the original data (colors), with the fixed position angle of 71.4$^\circ$ overplotted as a line { (the residuals are very similar for the case of CO/H$_2$ = 1)}.  The residuals in the integrated intensity map peak at roughly 16\% of the peak value of the observed moment zero map, but interestingly they display a departure from axisymmetry.  Along the southeast limb of the disk, the peak in the residuals is offset to the north of the disk major axis, whereas along the northwest limb of the disk the residual peak is offset to the south -- a geometry suggestive of spiral arms, a warp, a second disk with a different { position angle} \citep[like $\beta$ Pictoris][]{mat16}, or other non-axisymmetric structure in the disk.  { Running the fit with a variable position angle yields a best-fit PA of -77.9$^\circ$, which differs significantly from the continuum position angle (-71.4$^\circ$ $\pm$ 0.5$^\circ$) as well as the two best-fit position angle values obtained from fitting the integrated CO map with a 2-D Gaussian (-72.0$^\circ$ +/- 0.2$^\circ$) and from fitting the position angle of the two flux peaks in the integrated CO map (-70.0$^\circ$ +/- 0.5$^\circ$). While the position angle of the integrated CO map is statistically indistinguishable from that of the continuum map, the results of the channel map fit support a small but statistically significant deviation from axisymmetry in the CO map that manifests as a difference in position angle.}

{ An important point to consider is the optical depth of the CO emission from 49 Ceti.  The best-fit LIME models are at least marginally optically thin across a significant area of the disk for both classes of models: for the case of CO/H$_2$ = 10$^{-4}$ fewer than 0.4\% of the pixels with flux greater than 1\% of the peak value in the CO(3-2) model have $\tau > 1$, whereas for the case of CO/H$_2$ = 1 the optical depth is higher with $\sim20$\% of the pixels with flux $>1$\% of the peak exhibiting $\tau>1$ (in this latter case the optical depth rises above 10 across only 10\% of the disk area).  It is also possible to estimate the optical depth of the emission from the 3-D data cube.    If we assume LTE and} calculate the optical depth under that assumption from the location exhibiting the peak flux per beam per spectral resolution element in the CO(3-2) position-position-velocity channel maps (since this location presumably has the highest optical depth), following e.g. \citet{mat16}:
\begin{eqnarray}
\tau_\nu = \frac{h \nu}{4 \pi \Delta \nu} \left( \frac{x_l}{x_u} B_{lu} - B_{ul} \right) \frac{4 \pi d^2 F_{ul}}{h \nu A_{ul} \Delta A}
\end{eqnarray}
where h is Planck's constant, $\nu$ is the frequency of the spectral line, $\Delta \nu$ is the line width (1.7\,km\,s$^{-1}$ at the location of peak flux), $x_l$ and $x_u$ are the fractional populations in the lower and upper level of the transition, respectively, $B_{lu}$, $B_{ul}$, and $A_{ul}$ are the Einstein coefficients for the transition, $d$ is the distance to the star, $F_{ul}$ is the integrated flux density of the line (0.28\,Jy\,km\,s$^{-1}$ at the location of peak flux in the 3-D data cube, integrating across velocity space), and $\Delta A$ is the linear area of sky intersected by the emission along the line of sight (in this case, one synthesized beam ellipse with major and minor axis lengths of 0\farcs47 and 0\farcs32, respectively).  Assuming that the level populations are thermalized at a kinetic temperature of 40\,K (appropriate for the $\sim$100\,au distance of the emission peak from the star based on our two-line fit to the excitation temperature), this calculation yields a peak optical depth of 0.5.  While the calculation is sensitive to the assumed temperature, { even in the case where the excitation temperature corresponds to the 32\,K value indicated by the integrated line ratio the peak optical depth is only slightly higher, with a value of 0.66.  Departures from LTE can raise the value further, but the estimate is generally consistent with the output of the LIME models for the system.}  
%Finally, from a purely practical standpoint, the observation that the surface density of the CO(3-2) gas in the best-fit model increases with distance from the star further supports the assumption of low optical depth, since it is unlikely that the excitation temperature of the gas increases monotonically with distance from the central star (although it is possible that the excitation temperature could level out to a constant value with radius as the interstellar radiation field begins to dominate the excitation).  The surface brightness would therefore be tracing the product of temperature and surface density (implying optically thin emission), so that the approximately flat surface brightness with distance from the central star can be attributed to an increase in surface density with distance rather than temperature.  

The radial extent of the CO emission is also noticeably smaller than that of the dust continuum emission.  While it is difficult to compare directly the power-law surface density profile that we assumed for the dust with the similarity solution profile that we assumed for the gas, it is nevertheless clear that the observed outer radius of CO emission (220\,au when convolved with the beam, as we reported in Section~\ref{sec:co_obs}) is substantially smaller than the $\sim$310\,au outer radius of the best-fit models for the continuum emission.  In the context of our assumption of a CO/H$_2$ ratio of $10^{-4}$, the best-fit midplane density structure of the CO disk falls below the critical density of the CO(3-2) transition at a radius of $\sim$105\,au.  A substantially lower H$_2$ density would result in the density structure reaching this value at an even smaller radius.  This calculation is also affected by the departure from LTE and the uncertain kinetic vs. excitation temperatures in the disk.  

{ Comparison between the model with primordial abundances (CO/H$_2$ = 10$^{-4}$) and the model with low H$_2$ content (CO/H$_2$ = 1) results in parameters that obtain consistent results for stellar mass, inner radius, characteristic radius, inclination, and a scale height that is spatially unresolved (although the scale height factor is larger as expected for the higher mean molecular weight gas).  The disk mass of the low-H$_2$ model is lower, which is easily understood due to the lack of (invisible) H$_2$ that populated the primordial model and made up the bulk of the mass in that condition.  The surface density power law is steeper for the low-H$_2$ model; the reason is not entirely clear, but it may be related to the higher optical depth of the low-H$_2$ model which requires a more extreme density slope to generate sufficient flux contrast between the inner and outer disk (there is also likely to be a higher uncertainty on $\gamma$ in this more optically thick case).  The temperatures listed in Table~\ref{tab:best_fit_co} are the kinetic temperatures that LIME takes as its input; the unreasonably low value of 14\,K for the kinetic temperature of the low-H$_2$ model is probably not physical but may be related to the need to fine-tune the collision partner densities, which is beyond the scope of this paper's attempts to present a broad initial characterization of the gas and dust properties of the disk.  Nevertheless, the comparison between the two classes of models is instructive in revealing that the increasing surface density as a function of radius, along with the basic geometrical and kinematic properties of the system, is a result that is robust to large changes in the collision partner density in the disk. }

\begin{table*}
\caption{Best-fit parameters for parametric model fit to ALMA CO(3-2) and SMA CO(2-1) data}
\centering
\begin{tabular}{|l|cc|cc|c|}
\hline
Parameter & Best Fit & Range of Values & Best Fit & Range of Values & Finest Spacing \\
\hline
 & \multicolumn{2}{|c|}{CO/H$_2$ = 10$^{-4}$} & \multicolumn{2}{c|}{CO/H$_2$ = 1} &  \\
\hline
$M_*$ [M$_\sun$] & 2.1 & 1.5 to 3.0 & 2.1 & 1.5 to 3.0 & 0.1 \\
Log($M_\mathrm{disk}$ [M$_\sun$]) & -6.5 & -8 to -5& -9.0 & -10 to -7 & 0.2 \\
$R_c$ [au] & 140 & 50 to 200 & 140 & 50 to 200 & 20 \\
$R_\mathrm{in,gas}$ [au] & 20 & 10 to 100 & 20 & 10 to 100 & 10 \\
$\gamma$ & -0.5 & -2.0 to 1.6 & -1.8 & -2.0 to 1.6 & 0.2 \\
$T_{100}$ [K] & 40 & 10 to 100 & 14 & 10 to 100 & 2 \\
$i$ [$^\circ$] & 79.8 & 60 to 90 & 79.1 & 60 to 90 & 0.1 \\
$h_0$ & 0.75 & 0.2 to 1.2 & { 2.1} & 0.2 to { 3.0} & { 0.1} \\
\hline
%\multicolumn{4}{c}{CO/H$_2$ = 1} \\
%\hline
%$M_*$ [M$_\sun$] & 2.1 & 1.5 to 3.0 & 0.1 \\
%Log($M_\mathrm{disk}$ [M$_\sun$]) & -9.0 & -10 to -7 & 0.2 \\
%$R_c$ [au] & 140 & 50 to 200 & 20 \\
%$R_\mathrm{in,gas}$ [au] & 20 & 10 to 100 & 10 \\
%$\gamma$ & -1.8 & -2.0 to 1.5 & 0.2 \\
%$T_{100}$ [K] & 14 & 10 to 100 & 2 \\
%$i$ [$^\circ$] & 79.1 & 60 to 90 & 0.1 \\
%$h_0$ & 0.70 & 0.2 to 1.2 & 0.05 \\
\end{tabular}
\label{tab:best_fit_co}
\end{table*}

\section{Discussion}
\label{sec:discussion}

49 Ceti's unusual dust disk and its substantial reservoir of molecular gas given its advanced age are unexpected features that help to illuminate the late stages of circumstellar disk evolution.  Our analysis of the dust disk (Section~\ref{sec:analysis_cont}) requires a population of small grains that extends from $\sim$4\,au to $\sim$60\,au, and a population of larger ($\sim$1.6\,$\mu$m) grains that extends from $\sim$60\,au to $\sim$300\,au.  The surface density of the outer dust disk { (beyond $\sim100$\,au)} generally decreases with radius, although with a { marginally} statistically significant enhancement near a radius of $\sim$110\,au, suggesting a region of enhanced collision between planetesimals or a peak in the gas pressure where large grains collect.  Analysis of the CO emission (Section~\ref{sec:analysis_co}) suggests that the gas distribution differs substantially from that of the dust, displaying a surface density profile that increases with radius and is smaller in radial extent than the dust.  While $\sim$80\% of the CO(3-2) emission can be explained by an axisymmetric Keplerian disk, significant residuals at $\sim$20\% of the peak flux reveal a deviation from the dominant position angle of the gas and dust disk, suggestive of spiral arms or a warp in the gaseous material.  

\subsection{Comparison of Dust Surface Density Profile with ALMA-Resolved Debris Disks}

The resolution and sensitivity now available with ALMA are making it possible to characterize the radial surface density profile of the large grains in debris disks for the first time. Unlike the smaller, $\mu$m-size grains probed by optical and near-infrared observations of debris disks, the ALMA images are most likely primarily sensitive to large grains that trace the gravitational potential of the debris disk system and are not strongly affected by stellar radiation pressure.  They are therefore thought to trace the location of the parent planetesimal belts in the disk, and to provide an indication of the collision rate (which controls the dust production rate) between the parent planetesimals.  Thus far, only a handful of debris disks have had their radial surface density profiles { (i.e., whether the surface density increases or decreases with radius)} characterized in detail using high-sensitivity ALMA data.  Some disks, like that around $\beta$ Pic { \citep{den14}}, have been imaged in sufficient detail but are not sufficiently axisymmetric to permit the derivation of a meaningful dependence of surface density on radius.  { \citet{den14} do derive a nonparametric estimate for the surface density profile of the disk, which is comparable to the derived 49 Ceti profile in that it first increases and then decreases with radius, showing a local enhancement at a distance of $\sim$100\,au from the central star.}  Others, including Fomalhaut, HD 181327, 
$\eta$ Corvi, HD 21997, and a sample of the brightest debris disks in the Sco Cen region, are either intrinsically too radially narrow or have only been observed at too coarse an angular resolution { ($\Delta R / d \lesssim \theta$, where $\theta$ is the angular resolution of the data and $d$ is the distance to the system)} for the dependence of surface density on radius to be reliably measured \citep{bol12,moo13,whi16,lie16,mar16,mar17}.  For example, while the power law index $p$ describing the dependence of surface density on radius was included in models of the continuum emission from HD 21997, the uncertainty on the measured value of $-2.4\pm4.5$ is too large to meaningfully distinguish between rising and falling surface density as a function of radius in that system.  The other systems that have been observed to date that are intrinsically radially broad enough, axisymmetric enough, and have been observed at sufficiently high angular resolution to permit a detailed characterization of their dependence of surface density on radius include AU Mic and HD~107146.

The M-type star AU Mic, a member of the $\beta$ Pic moving group with an age of $\sim$20\,Myr, hosts an edge-on debris disk that displays no CO emission { (Hughes et al.~in prep)}.  Millimeter-wavelength observations with the SMA by \citet{wil12} and with ALMA by \citet{mac13} { show it to be consistent with a dust surface density increasing } steeply with radius as $\Sigma \propto r^{2.8}$ between radii of 10-40\,au before being abruptly truncated.  This surface density profile is consistent with predictions for a self-stirred disk with ongoing planet formation \citep{ken10}, but the timescale required to initiate the collisional cascade is much longer than its age \citep[see][]{ken08}.   \citet{str06} suggest that a birth ring exists at a radius of $\sim$40\,au, in which grains of all sizes are created through collisions between parent planetesimal bodies.  The smaller grains, highly susceptible to stellar winds and radiation pressure, are blown out on eccentric orbits forming an extended halo visible in scattered light \citep{kal04,kri05,gra07,sch15}, { and exhibit time-dependent behavior \citep{boc15}}.  

The 100\,Myr-old G star HD~107146 hosts a debris disk that has been previously spatially resolved at millimeter wavelengths with CARMA \citep{cor09} and the SMA \citep{hug11}.  Cycle 0 observations with ALMA at an angular resolution of 1\,arcsec have spatially resolved the radial surface density profile \citep{ric15}.  The ALMA continuum emission between radii of 30 and 150\,au show an overall trend of increasing surface density with radius, similar to AU Mic.  However, mirroring the surface density enhancement that we have measured in the 49 Ceti disk, there is evidence for a local {\it decrement} of surface density at a radius near 80\,au.  The simplest explanation for the clearing in the dust disk is that a planet of a few $M_\earth$ has coalesced and cleared a nearly circular orbit at a distance of $\sim$80\,au from the central star.  A planet on a high-eccentricity orbit with smaller semimajor axis has also been proposed as the cause \citep{pea15}.  The increasing surface density of the outer disk is only marginally consistent with a self-stirring scenario, requiring an initial mass surface density roughly ten times that of the minimum mass Solar nebula to initiate stirring on a timescale comparable to the age of the system.  

49 Ceti's best-fit surface density power law of $p = 1.29$ in the single power law model is comparable to the typical surface density power law of $p \sim 1$ observed for optically thick protoplanetary dust disks in nearby young star forming regions \citep[see, e.g.,][]{and07}, perhaps suggesting that 49 Ceti's dust disk is a scaled-down (and hollowed-out) remnant of the initial protoplanetary disk.  In this scenario, the local enhancement at a radius of $\sim110$\,au might correspond to the outer boundary of a self-stirred region, with the outer component perhaps composed of a significant fraction of primordial grains. 
%The power-law profile is also consistent with the radial surface brightness profile of $\Sigma \propto R^{-1.5}$ that has been predicted for radiation pressure-affected grains in collisionally dominated debris disks outside the birth ring \citep{str06,the08}; the power law index is largely independent of gas content, composition, or temperature \citep{kri09}. However, the radiation pressure-affected grains are substantially smaller than those generally thought to produce the bulk of the milimeter-wave emission observed by ALMA.  This threshold is typically predicted to occur at a ratio of radiation pressure to gravity $\beta \sim 0.05$, which corresponds to grain sizes of $\lesssim$32\,$\mu$m in the 49 Ceti system for a stellar luminosity of 15.5$L_\sun$ \citep{rob13}.  Typical assumptions about the underlying grain size distribution lead us to believe that the grain sizes probed are comparable to the wavelength of observation, i.e., that the ALMA observations probe grains close to 1\,mm in size.  However, if the grain size distribution were such that smaller radiation pressure-sensitive grains dominated the opacity at millimeter wavelengths, it could account for the observed surface brightness profile in the 49 Ceti disk.  Such a scenario would require that the grain size distribution be markedly different for the { gas-bearing} 49 Ceti disk than the gas-poor disks around AU~Mic and HD~107146 to explain the starkly different surface brightness profiles of gas observed at comparable wavelengths.  

Whatever its origin, the decreasing surface density profile in the outer regions of the 49 Ceti dust disk stands in sharp contrast to the increasing surface density profiles of AU Mic and HD~107146, which are consistent with theoretical predictions for self-stirred debris disks \citep[e.g.,][]{ken02,ken10}.  While only three debris disks have yet been observed at millimeter wavelengths with sufficient sensitivity and angular resolution to provide reliable constraints on the radial surface brightness profile of continuum emission, the results so far are suggestive that 49 Ceti's { gas-bearing} dust disk may be undergoing a fundamentally different physical process than that of the gas-poor debris disks around AU Mic and HD 107146, resulting in the qualitatively different surface density profiles.  With the new { gas-bearing} debris disks recently identified by \citet{moo15} and \citet{lie16}, it should soon be possible to determine whether or not this trend persists for a larger sample of objects.  

\subsection{Gas Surface Density Profile, and Interactions Between Gas and Dust}

A striking feature of the best-fit CO disk structure is that the surface density increases with distance from the central star (Section~\ref{sec:analysis_co} and Table~\ref{tab:best_fit_co}), independent of the density of the collision partner.  { The difference is statistically significant; when the surface density power law $\gamma$ is fixed at zero while the other parameters are varied, the best-fit model has a $\chi^2$ value that exceeds the best-fit model with negative gamma by a difference of $\Delta\chi^2 = 1631$, which represents $>10\sigma$ significance according to the AIC.}  This result, along with the central gap in the CO disk, is a clear departure from the typical surface density structure of protoplanetary disks around pre-main sequence stars, which exhibit surface density profiles that decrease with distance from the central star -- very consistently in continuum emission, but also in the few cases for which the gas surface density profile has been measured using low optical depth molecular lines \citep[e.g.,][]{pie05,pan08,qi11,fla15,sch16}.  While increasing gas surface density profiles have been observed within the central cavity of transition disks \citep{car14,vdmar16}, the same disks exhibited evidence for decreasing surface density of gas in the outer disk.  It should be noted that the quantity directly probed by the data is the surface brightness profile, $p + q = 0$ for our assumed temperature power law of $q = 0.5$, but nonetheless the surface brightness is approximately constant with distance from the central star, in stark contrast with previous observations of gas disks.  

At the time of writing we are not aware of any theoretical papers that consider or predict gas surface density profiles in debris disks that increase with radius, making the interpretation of this result challenging.  In a steady-state scenario in which the gas is produced and destroyed at a constant rate, the CO volume density at a given radius will depend on the gas production rate and the destruction rate.  { In the case of} negligible shielding of CO from photodissociation, the photodissociation rate should be proportional to the UV flux.  In regions of the disk where the stellar contribution dominates over the interstellar component of the UV radiation field, the UV flux and therefore the photodissociation rate would dilute with radius as $R^{-2}$.  If the gas production rate follows a power law profile described by $R^{-x}$, then the final radial density distribution of CO would be described by a power law of $R^{2-x}$.  In this scenario, our best-fit radial volume density distribution of CO in the midplane $\rho_0 \propto \frac{\Sigma(R)}{h(R)} \propto R^{-0.75}$ would imply a gas production rate (in units of surface density) that varies with radius as $R^{-1.5}$, comparable to the dust surface density in the 49 Ceti system and to a typical protoplanetary disk surface density profile.  \citep[This only applies to the case of negligible shielding.  In the presence of sufficient quantities of neutral carbon shielding might change the dependence of photodissociation rate on radius; see][]{kra16}. { If} the dynamical timescale ($\sim$130\,yr at a radius of 100\,au) is comparable to or shorter than the CO photodissociation timescale, { then} the produced CO might be plausibly axisymmetrically distributed throughout the disk by Keplerian shear.  

It is also important to consider whether the quantity of gas in the system is sufficient to cause noticeable effects on the dust due to interactions between the two components.  There have been several theoretical investigations of the effects of gas dynamics on the solid component of debris disks.  The pioneering work of \citet{tak01} predicted that for A stars surrounded by a total gas mass ranging from a fraction to dozens of Earth masses (within which range our observations place the dust mass of the 49 Ceti disk, with the exact value depending on the assumed CO abundance relative to H$_2$), small dust grains (with sizes $\sim$10-200\,$\mu$m) will start to orbit at sub-Keplerian speeds due to Poynting-Robertson drag and subsequently experience a tailwind from the (slightly less sub-Keplerian) gas that can eject them out to orbital radii beyond the edge of the gas disk.  This dust migration can result in either a ring just beyond the outer edge of the gas disk or a bimodal radial dust distribution.  In related work by \citet{kla05}, they demonstrate a general result that gaps and sharp edges can form in debris dust due to instabilities within optically thin gas disks where the gas temperature is an increasing function of the dust surface density.  Subsequent studies by \citet{her07} and \citet{bes07} added photophoresis and photoelectric effects, respectively, to the calculation.  \citet{her07} demonstrated that photophoresis generally increases the stability radius for a grain of a given size (as does a higher gas density, a lower absolute value of the temperature power law index $q$, or higher stellar luminosity).  They also point out that the outer ring-like structure described by \citet{tak01} might not be so sharp in a disk with a less abruptly truncated outer edge.  This might help to explain the more gradual decrease in dust density beyond the outer edge of the gas disk that we observe in the 49 Ceti system.  The observation that the radial extent of the CO disk is smaller than that of the dust continuum emission is both in sharp contrast with observations of protoplanetary disks \citep[e.g.,][]{and12} and  consistent with an emerging trend in the radial profiles of other { gas-bearing} debris disks as well.  \citet{fla16} and \citet{whi16} report observations showing that the CO emission from the HD 141569 disk is confined to within the smaller of the two scattered light rings, while \citet{lie16} show that for a sample of 23 debris disks in the Sco-Cen region, the three { gas-bearing} disks in the sample exhibit preferentially large radii for their dust continuum emission.  
However, these effects would require that the dust grain size distribution in 49 Ceti be such that the dominant source of opacity in the 850\,$\mu$m image of the disk be grains smaller than $\sim$200\,$\mu$m that are sensitive to Poynting-Robertson drag, which may pose additional problems for describing the surface brightness distribution \citep[e.g.,][]{the05} and is likely inconsistent with the observed millimeter spectral index of the system.  

\citet{bes07} point out that in disks where the gas-to-dust mass ratio is $\gtrsim$1 and the dust mass is at least $\sim0.005$\,M$_\earth$, photoelectric heating effects will enhance any overdensity of dust grains (e.g., due to a recent catastrophic collision).  This instability tends to create narrow rings of enhanced dust density, where the ring location is not sensitive to the radial profile of the gas disk.  A similar mechanism is also described with more detailed simulations by \citet{lyr13}, who predict a clumping instability that organizes dust into narrow, eccentric rings within { gas-bearing} debris belts (again, in systems with gas-to-dust mass ratios $\gtrsim1$).  Such an instability might help to explain the local enhancement of dust surface density at a radius of $\sim$110\,au in the 49 Ceti disk, perhaps indicating the recent production of dust through a destructive collision between planetesimals.  Although the gas-to-dust mass ratio in the 49 Ceti system is not well constrained, it seems plausible that it might lie in the $\gtrsim1$ regime required by the \citet{bes07} and \citet{lyr13} theories.  The assumption of an ISM-like CO/H$_2$ ratio of $10^{-4}$ yields a global gas-to-dust mass ratio of 6 for the double power-law model, while an assumption that CO provides the dominant contribution to the gas mass results in a gas-to-dust mass ratio of 0.008.  It is also important to consider that the local gas-to-dust mass ratio might also exceed these values  (due to unresolved local vertical or radial density variations), although it is difficult to determine precisely where this might occur since the spatial resolution of our observations is insufficient to resolve the vertical structure of the gas disk.  

One puzzling feature of the dust emission that might be explained by dynamical interactions between gas and dust is the apparent presence of a population of dust grains substantially smaller than the blowout size for the 49 Ceti system, previously remarked upon by \citet{wah07}.  The presence of an inner disk of small ($\sim$0.1\,$\mu$m) dust grains is echoed by a recent survey for gas emission in 10\,Myr-old debris disks in the Sco-Cen region, which demonstrated that { gas-bearing} debris disks appear more likely than gas-poor debris disks to demonstrate evidence for a population of grains smaller than the blowout size, based on a simultaneous analysis of their SEDs and spatially resolved visibilities \citep{lie16}.  In that survey, which included three { gas-bearing} debris disks, two of the three disks with evidence for a population of dust below the blowout size were { gas-bearing}, and the third { gas-bearing} debris disk exhibited a characteristic grain size only slightly larger than the blowout size.  It is possible that drag forces from the substantial quantity of molecular gas in these systems is preventing the removal of small grains by radiation pressure.  It is also possible that these systems have undergone more recent, active collisions that have generated both the gas and the small dust grains that have not yet been removed from the system; however as \citet{lie16} observe, such a scenario would tend to predict that the high gas masses would be accompanied by concomitantly high dust masses, yet the survey reveals no obvious bias towards high dust mass among the { gas-bearing} debris disks in the sample.  

The lack of a clear silicate emission feature in the {\it Spitzer} IRS spectrum of 49 Ceti is somewhat puzzling in the context of the small grain population.  There are at least two plausible explanations for its absence.  First, silicate features are most readily apparent when emitting grains are substantially smaller than $\sim$1\,$\mu$m \citep{pap83}, although \citet{nat07} show that a 10\,$\mu$m silicate feature is also apparent in the emission of 200-600\,K grains smaller than a few microns.  Our best fit models suggest an inner radius of $\sim$5\,au for the inner dust component, with a substantial amount of degeneracy in the absence of spatially resolved data.  At this location, or if the true location of the warm component is somewhat further from the central star, the grains may be not quite hot enough to excite a silicate emission feature.  Alternatively, it is possible that the grains are predominantly carbonaceous in nature rather than silicate, which would also alter the density structure of the disk.  This scenario is supported by the { low albedo} of 49 Ceti in previous {\it HST} NICMOS observations searching for scattered light in the outer disk \citep[][]{wei99,cho17}, and by the evidence for carbon overabundance in the atomic gas disk from line-of-sight UV absorption spectroscopy \citep{rob14}.  

The ALMA observations also allow us to place the best constraints to date on the vertical structure of the gas disk.  We find a best-fit scale height factor $h_0$ of 0.75 for the primordial abundance scenario, indicating that the observed scale height is substantially smaller than that expected for a typical H$_2$-rich disk in hydrostatic equilibrium.  A scale height factor less than one cannot be easily explained by the vertically isothermal approximation (any appreciable optical depth would cause the disk surface to be hotter than the midplane, puffing up the disk beyond the height expected from the midplane temperature), nor by a scenario in which the kinetic temperature exceeds the excitation temperature (if the kinetic temperature is significantly hotter than the excitation temperature, we would expect a more, not less, vertically extended disk).  A departure from LTE could be responsible for the difference only if the kinetic temperature were lower than the excitation temperature, e.g., in the case of strong UV pumping in a very low density gas that is not thermalized by collisions with atomic or molecular hydrogen.  The scale height factor should also be considered an upper limit, since the best-fit disk structure exhibits a scale height approximately equal to the spatial resolution at the largest radii of the gas disk.  The simplest explanation for a scale height smaller than that predicted { for hydrostatic equilibrium in the primordial abundance scenario} is that the primary constituent of the gas disk is not molecular hydrogen.  Since the scale height is proportional to the sound speed, which scales as $1/\sqrt{\mu}$, if the gas mass is dominated by CO instead of H$_2$, we would expect a scale height factor of $\sqrt{2.37/28} = 0.3$, which is consistent with our { upper limit for the primordial abundance scenario}, and with the best-fit value of 2.1 derived for the low-H$_2$ scenario (which should also be considered an upper limit since it too is spatially unresolved).  A small scale height is interesting because it increases the potential for CO self-shielding along the disk midplane and might play a role in enabling longer lifetimes for volatile-rich molecular gas disks despite the energetic radiation from the central A star.  A lower H$_2$ gas mass would also indicate that dynamical interactions between gas and dust are probably not significant, and would provide some support for a second-generation origin for the CO. Higher-resolution observations of the 49 Ceti disk are necessary to better constrain the vertical structure of the gas disk, and thereby understand the potential for CO self-shielding.  

\subsection{Dust Stirring Mechanisms}

Since it is unclear whether or not there is sufficient gas to affect dust dynamics in the 49 Ceti system, we consider dust stirring mechanisms in the context of predictions for typical gas-poor debris disks, with the caveat that depending on the composition of the gas it might have an effect on the apparent properties of the dust disk.  The two main mechanisms proposed for generating dust in debris disk systems are self-stirring, in which ensembles of Pluto-size planetesimals excite eccentricity \citep[e.g.,][]{ken04}, and stirring by a planetary-mass object \citep[e.g.][]{mus09}.  { Other possibilities include stochastic collisions or steady-state collisional evolution of pre-stirred bodies \citep[for an overview, see][]{wya08}.}

The self-stirring mechanism begins to operate with the formation of the first objects of $\sim1000$\,km in scale.  \citet{ken08} derive a timescale of
\begin{eqnarray}
t \simeq 145~ x_m^{-1.15} \left( \frac{R}{80\mathrm{au}} \right)^3 \left( \frac{2 M_\sun}{M_*} \right)^{3/2}~\mathrm{[Myr]}
\end{eqnarray}
for the formation of objects sufficiently large to excite self-sirring, where $R$ is the semi-major axis of the dust belt, $M_*$ is the mass of the central star, and $x_m$ is a scaling factor used to describe the initial surface density of the protoplanetary disk $\Sigma_i$:
\begin{eqnarray}
\Sigma_i(a) = \Sigma_0 (M_*)~ x_m \left( \frac{R}{R_0} \right)^{-3/2}~\mathrm{[g\,cm}^{-2}]
\end{eqnarray}
The typically assumed reference surface density is the minimum mass solar nebula \citep[MMSN;][]{wei77}, for which $\Sigma_0 \simeq 0.18$\,g\,cm$^{-2}$ at $R_0 = 30$\,au.  \citet{ken08} assume that this surface density scales linearly with the mass of the central star, such that $\Sigma_\mathrm{30~au}(M_*) \simeq 0.18~\left(\frac{M_*}{M_\sun}\right)$\,g\,cm$^{-2}$.  Solving for the scaling factor needed to produce a ring at 110\,au within the 40\,Myr age of 49 Ceti yields a factor $x_m \simeq 8$, or an initial surface density of solids around 49 Ceti of roughly 16$\times$ the MMSN.  \citet{mus09} demonstrate that $x_m$ must be $<10$ to avoid gravitational instability for a canonical 100:1 gas-to-dust mass ratio, implying that 49 Ceti's initial protoplanetary disk disk would need to be close to the point of gravitational instability in order for self-stirring to be a viable mechanism to produce the observed surface density enhancement.  

\begin{figure}[t!]
\centering
\includegraphics[scale=0.45]{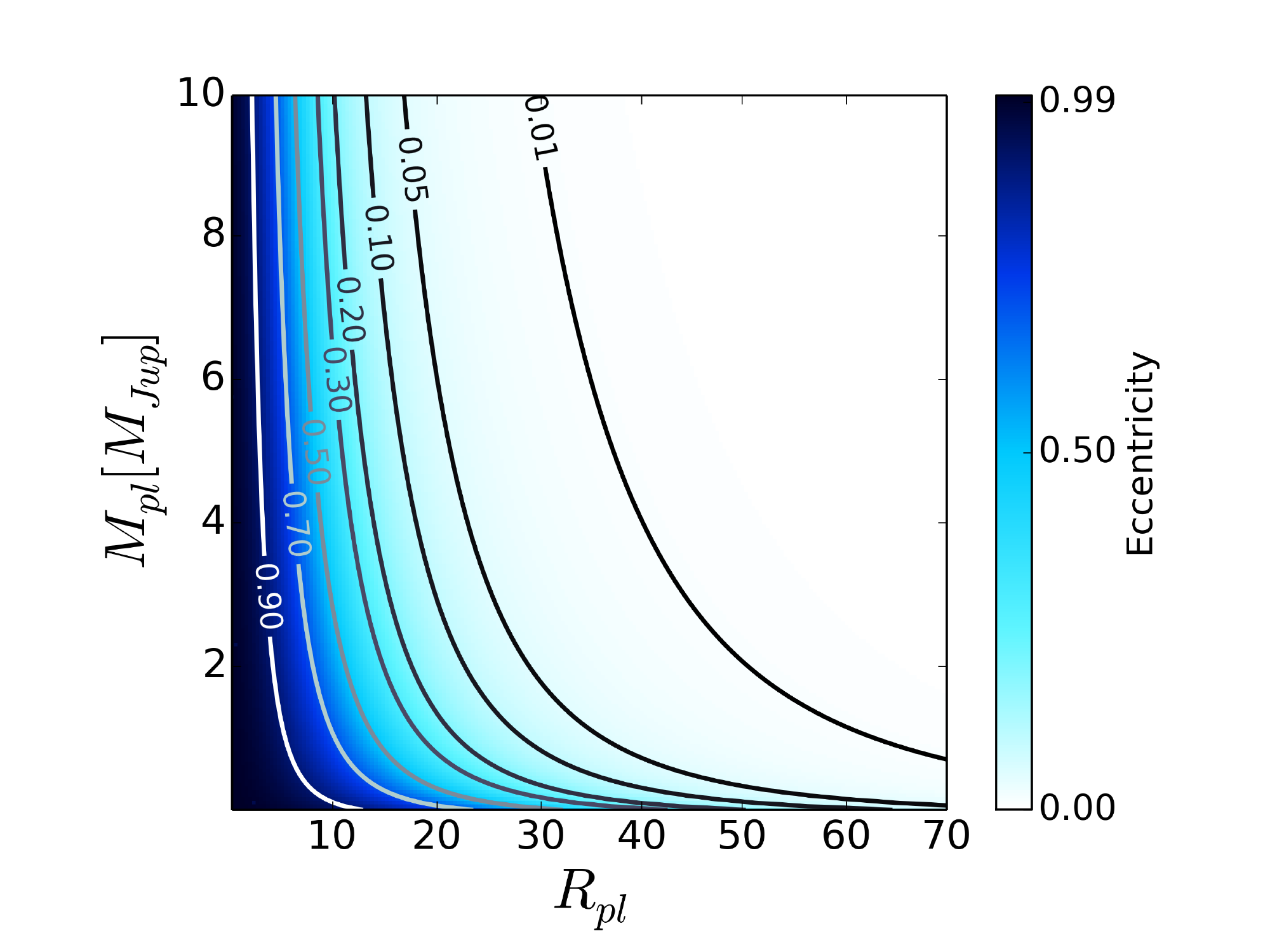}
\caption{
The eccentricity in $R_{pl}$-$M_{pl}$ parameter space needed to stir the observed planetesimal belt at $\sim$110\,au in less than the 40\,Myr age of the 49 Ceti system.  The contour lines correspond to $e_{pl}$ = 0.01, 0.05, 0.2, 0.3, 0.5, 0.7, and 0.9.  
}
\label{fig:planet}
\end{figure}

The ability of a planet to excite dust production through its gravitational influence generally depends on its mass ($M_{pl}$), orbital eccentricity ($e_{pl}$) and semimajor axis ($R_{pl}$), as well as the mass of the central star ($M_*$).  The required relative excitation velocity for destructive impacts between equal-size bodies in the outer disk, where weak conglomerations of ices tend to dominate, is derived by \citet{mus09} as
\begin{eqnarray}
\resizebox{0.4\textwidth}{!}{
$v_{rel}(a) = \left[ 0.8 \left(\frac{a}{\mathrm{80~m}} \right)^{-0.33} + 0.2 \left(\frac{a}{\mathrm{80~m}} \right)^{1.2} \right]^{0.83}~\mathrm{[m~s}^{-1}]$
}
\end{eqnarray}
where $a$ is the size of the body.  Destructive collisions between grains of size $a \sim 1-10$\,cm can produce the 100\,$\mu$m-1\,mm particles that are thought to dominate the emission at the wavelength of the ALMA image, although larger bodies such as km-size comets may also be the parent bodies of the observed dust (and gas) -- either parent body requires collisional velocities of order 5-10\,m\,s$^{-1}$ to result in destructive collisions.  For an internal perturber, the semimajor axis $R^*$ within which a planet can excite grains of radius $a$ to $v_{rel}$ is derived by \citet{mus09} as
\begin{eqnarray}
\resizebox{0.4\textwidth}{!}{
$R^*(a) = 3.8~\left(\frac{e_{pl}}{0.1}\right)^{2/3} \left(\frac{M_*}{1~M_\sun}\right)^{1/3} \left(\frac{R_{pl}}{\mathrm{1~au}}\right)^{2/3} \left(\frac{v_{rel}(a)}{\mathrm{1~km~s}^{-1}}\right)^{-2/3}~\mathrm{[au]}$
}
\end{eqnarray}
while the timescale for stirring to occur is given by
\begin{eqnarray}
\resizebox{0.4\textwidth}{!}{
$t \sim 1.53\times10^3 \frac{(1-e_{pl}^2)^{3/2}}{e_{pl}} \left(\frac{R}{\mathrm{10~au}}\right)^{9/2} \left(\frac{M_*}{1~M_\sun}\right)^{-1} \left( \frac{M_\mathrm{pl}}{M_\sun}\right)^{-1} \left(\frac{R_{pl}}{\mathrm{1~au}}\right)^{-3}~\mathrm{[yr]}.$
}
\end{eqnarray} 
Figure~\ref{fig:planet} displays the eccentricity required for a planet of semi-major axis $R_{pl}$ and mass $M_{pl}$ to stir the planetesimal belt at 110\,au in less than 49 Ceti's age of 40\,Myr.  For example, a Jupiter-like planet with $e_{pl} \sim 0.05$ and $M_{pl} \sim 10^{-3}M_\sun$ located at a distance of 40\,au from the central star could excite planetesimals to collisional velocities within the dust ring in less than the age of the system.   

\subsection{Origin of the Gas}

49 Ceti joins a handful of { gas-bearing} debris disks that have been imaged with ALMA, and therefore have high-quality images of both the molecular gas disk and thermal dust continuum emission: the others are HD 21997 \citep{kos13,moo13}, $\beta$ Pictoris \citep{den14}, HD 141569 \citep{whi16}, HD 181327 \citep{mar16}, HIP~76310, HIP~84881, and HIP~73145 \citep{lie16}.  Among these systems, 49 Ceti is the first with sufficiently { axisymmetric} structure and sufficiently high-resolution and high-sensitivity observations that the surface density profile of both gas and dust can be studied in detail.  However, previous observations have provided clues to the origin of the gas: the investigations center around whether the gas is remnant primordial material that has somehow managed to survive despite the destructive influence of high-energy radiation from the central star, or whether the gas is second-generation material that was previously incorporated into planetary embryos, planetesimals, or the icy mantles of dust grains.  

Structurally, $\beta$ Pictoris is unique among the disks imaged in exhibiting a very strong asymmetry of both gas and dust emission, with a large clump of emission appearing along one limb of the disk (although since the $\beta$ Pic disk is viewed nearly edge-on to the line of sight, it is impossible to uniquely deproject the radial structure of the feature).  The structural asymmetries, along with the surprising presence of a number of atomic species revealed by absorption spectroscopy, provided strong hints that the gas and dust were likely of secondary origin, resulting from a recent collision between Mars-size bodies or collisional evaporation of icy planetesimals at resonances induced by the presence of a giant planet \citep{den14}.  A recent analysis of { optically thin} CO { line ratios} from the system determines conclusively that the gas must be of second-generation origin, thanks to a non-LTE analysis demonstrating that the H$_2$ densities must be too low to shield CO from photodissociation, { and disfavors the collision of Mars-size bodies as the origin of the gas} \citep{mat16}.  However, it is not clear to what extent $\beta$ Pictoris is typical or atypical compared to other { gas-bearing} debris disks, since all the other systems have shown at most minor deviations from axisymmetry { at least at the $\sim10$\% level enabled by the contrast ratios of current observations} and are predominantly characterized by ring- or disk-like structures in both gas and dust.  

Among the more regular rings and disks observed around the other stars, arguments about the origin of the gas tend to focus on the following characteristics:

{\it Relative location and structure of gas and dust --} An initial clue to the physical mechanisms underlying the origin of gas and dust comes from their relative locations in the disk.  If both the gas and dust are second-generation, produced in similar events involving collisional destruction of planetesimals, then one might expect their physical locations in the disk to be similar.  If, on the other hand, the gas is primordial while the dust is second-generation, the gas might follow the radial profile of the primordial protoplanetary disk while the dust exhibits ring-like structures more characteristic of an archetypal debris disk.  Muddying the waters a bit, both configurations -- and in 49 Ceti, a configuration that does not obviously match either scenario -- have already been observed even in the small sample of disks with both thermal continuum and molecular gas emission images.  \citet{mar16} note that the CO emission, while faint and difficult to image, appears to follow the ring-like structure observed in dust continuum emission in the disk around HD 181327, and cite this co-location as potential evidence for a second-generation origin.  By contrast, \citet{kos13} observe a far more radially extended CO configuration than the dust continuum ring observed in the HD 21997 disk, and cite this difference as evidence that the gas may be primordial while the dust is second-generation in that system.  49 Ceti's gas and dust configurations exhibit different radial dependences as well as different inner and outer extents, perhaps suggesting that their physical origins differ as well -- although it is also suggestive to consider the possibility that the underlying gas production rate is diluted by the $R^{-2}$ UV flux, implying that the underlying surface density profile might in fact match that of the dust disk -- and that of a typical protoplanetary disk rather than a typical self-stirred debris disk.  Similarly high-resolution and high-sensitivity observations of the other { gas-bearing }debris disks will provide a more thorough understanding of the relative location of gas and dust in the known { gas-bearing} debris disk systems.  

{\it Molecular gas densities and lifetimes -- } Another cornerstone of the discussion has been the true quantity of molecular gas in the disk, which centers around the question of the gas composition, and particularly the CO-to-H$_2$ abundance and whether or not it differs from the primordial value of $10^{-4}$.  Interstellar abundances of $H_2$ might be capable of shielding CO from destruction by the harsh stellar UV radiation fields, implying that primordial CO might be able to survive to the 10-40\,Myr ages of the stars surrounded by substantial molecular gas disks.  In the absence of a substantial H$_2$ population, by contrast, the CO photodissociation timescales are quite rapid in these systems, of order hundreds to thousands of years depending on the details of the spatial distribution of CO \citep{vis09}.  Circumstantial evidence in several systems has implied that the H$_2$ content of the disks may be quite low, including low excitation temperatures derived from multi-transition and/or { multi-transition} analysis of CO that imply a lack of collision partners for the CO molecules \citep{kos13,fla16,mat16}.  There is also more conclusive evidence of a non-primordial CO/H$_2$ ratio in both $\beta$ Pic and AU Mic \citep{lec01,rob05}.  While the excitation temperature in the 49 Ceti disk is not obviously lower than the kinetic temperature, the spatially unresolved vertical structure in this nearly edge-on system provides circumstantial evidence of a mean molecular weight substantially higher than the ISM value of $\sim2.37$, lending support to the non-ISM CO/H$_2$ abundance and therefore the second-generation origin of material in the disk.  While the apparent trend of dust disks larger than their CO counterparts might be neatly explained by the presence of gas in sufficient quantities to create the tail-wind effect predicted by \citet{tak01}, this would also require that the dust grain size distribution be sufficiently weighted towards small ($<200$\,$\mu$m) grains that they dominate the opacity even at millimeter wavelengths.  

{\it Required production rates for secondary material --} One of the most significant challenges to the second-generation gas disks scenario is the question of whether it is plausible to sustain collision rates high enough to maintain the gas disk for a substantial fraction of the star's lifetime.  Standard calculations based on the rapid photodissociation timescales of hundreds to thousands of years suggest that it would be necessary to evaporate a large (Hale-Bopp-size) comet every few minutes to sustain the necessary gas production rates \citep{zuc12,kos13,den14,whi16}.  { Whether or not such collision rates are reasonable given the likely reservoirs of cometary material populating the collisional cascade is still a subject of discussion within the literature.  \citet{kos13} argue that the rate is at best marginally consistent with expected reservoir, while \citet{zuc12} argue that such a high cometary collision rate is consistent with the high dust luminosity of the HD 21997 system.  \citet{den14} calculate the collision rate without commenting on its plausibility.  \citet{whi16} calculate that there is potentially sufficient cometary material in the HD 141569 disk to produce the observed CO flux, although significant shielding is required to maintain the observed abundance for a significant length of time. Some disks do exhibit demonstrable consistency with the mass loss rate produced by a steady-state collisional cascade based on Solar System cometary abundances \citep{mat15,mat16,mar16}.} Chance collisions of larger, Mars-size bodies in the disk have the potential to release larger quantities of CO, but such statistically rare events are unlikely to explain the population of ``typical'' axisymmetric { gas-bearing} debris disks given observation that approximately half of intermediate-mass debris disk host stars with ages of $\sim$10\,Myr harbor { gas-bearing} debris disks \citep{lie16}.  A recent model suggests that evaporation from grains in the collisional cascade may be able to explain the quantity of gas observed in the 49 Ceti system (Kral et al. in prep), although this will need to be reconciled with the lack of correlation between dust and gas mass in debris disks observed by \citet{lie16}.  The best unexplored avenue for disentangling a cometary from a primordial gas disk origin is by examining the chemical composition of the molecular gas component.  While HCN seemed a promising molecule due to its relatively high abundance both in typical protoplanetary disks \citep{obe10} and its order-of-magnitude higher abundance relative to CO in Solar System comets \citep{biv02}, unfortunately the search for HCN in the current study of 49 Ceti was fruitless.   It is difficult to predict how the composition of second-generation material produced through collisions at large distances from the star might compare to that observed for photodesorption of ices on comets in the inner Solar System.  However other molecules that are not as readily photodissociated may be more likely to yield useful chemical insights that can distinguish the composition of primordial gas from that of comet-like material. 

At the moment, the most promising future avenues for disentangling the origin of the gas in debris disks would appear to be a non-LTE { multi-line} CO analysis of a larger sample of disks, along the lines of that presented in \citet{mat16}, perhaps combined with a chemical search for targeted molecular lines that might help to distinguish the composition of protoplanetary gaseous material from that of cometary material.  Combining these insights with measurements of the detailed spatial distribution of gas and dust in these systems will yield insight into the physical origins of the constituent material.  

\section{Summary and Conclusions}
\label{sec:conclusions}

We have presented ALMA imaging of the 49 Ceti system that resolves the radial surface density distribution of molecular gas and dust continuum emission.  This is also the first { gas-bearing} debris disk that is sufficiently axisymmetric and observed with sufficiently high sensitivity and resolution to characterize { changes in the surface density} of gas and dust { as a function of distance from the central star}.  

We derive a surface density distribution of millimeter dust emission that generally decreases with distance from the central star between radii of { $\sim$100} and $\sim$310\,au.  We model the spatially resolved ALMA continuum visibilities in tandem with the spatially unresolved broad-band SED and derive evidence for a two-disk structure, with an inner disk of small grains that is not detected in the millimeter continuum emission but whose properties are consistent with previous mid-IR imaging of the 49 Ceti system by \citet{wah07}.  We investigate three different classes of models for the spatial distribution of dust in the outer disk, ultimately demonstrating that while a single power-law surface brightness model provides an adequate fit to the data, a { marginally} statistically significant improvement is provided by models that include a surface density enhancement at a distance of $\sim$110 au from the central star.  The current data cannot distinguish between a double power-law profile that increases in surface density between $\sim$60 and $\sim$100\,au and decreases from $\sim$100 to $\sim$310\,au, and a single narrow ring at a distance of $\sim$110\,au from the central star superimposed on a broad power-law disk.  

The surface density distribution of CO is well described by a similarity solution model where density increases with radius between $\sim$20 and $\sim$220\,au from the central star.  Such a regular, axisymmetric disk in Keplerian rotation about the central star reproduces approximately 80\% of the flux in the zeroth moment map of CO(3-2) emission, while the residuals at roughly 20\% of the peak flux exhibit a non-axisymmetric structure suggestive of spiral arms or a warp in the gas disk.  There is also tentative evidence that the gas disk exhibits a scale height smaller than that predicted for the standard assumptions of hydrostatic equilibrium for a disk with an interstellar CO/H$_2$ abundance of 10$^{-4}$, which is consistent with previous non-detection of CO absorption features along the line of sight to the star \citep{rob14}.  The smaller than expected scale height perhaps indicates that the mean molecular weight of the gas is larger than ISM abundances would predict due to substantial depletion of H$_2$, { and is consistent with the upper limit of 2.1 times the hydrostatic equilibrium scale height from the best-fit model with low H$_2$ content}.  The small scale height therefore provides circumstantial evidence in support of a second-generation origin of the molecular material, although the small scale height could also help to increase CO self-shielding and thereby increase the expected lifetime of CO molecules in the disk.  

The surface density profile of the dust differs substantially from those of the more typically gas-poor debris disks around AU Mic and HD 107146 that have been previously observed with ALMA, which exhibit radial profiles that increase with distance from the star.  Likewise, the increase of gas surface density with distance from the central star is { very unusual for a circumstellar disk, and seems to persist even in the case of higher optical depth in a low-H$_2$ composition scenario}.  Efforts to similarly characterize other known { gas-bearing} debris disks at high spatial resolution and sensitivity would reveal whether these characteristics are unique to 49 Ceti, or are common to { gas-bearing} debris disks and therefore indicative of the underlying physical origin of the gas and dust reservoirs.  Future observations that would improve our understanding of the origin of gas include a { multi-line} non-LTE analysis of { optically thin} CO emission in a sample of known axisymmetric { gas-bearing} debris disks, and a survey to reveal the molecular chemistry of the gas in the 49 Ceti disk.  

\acknowledgments

The authors wish to thank Angelo Ricarte for his contributions to the code base, and Eugene Chiang and Quentin Kral for helpful conversations that improved the manuscript.  A.M.H., J.L.-S., and K.M.F. are supported by NSF grant AST-1412647.  C.~D. was sponsored by a NASA CT Space Grant Undergraduate Research Fellowship and a Wesleyan Research in the Sciences fellowship.  The work of A.K., A.M., and P.A. was sponsored by the Momentum grant of the MTA CSFK Lend\"ulet Disk Research Group.  { A.M. acknowledges support from the Bolyai Research Fellowship of
the Hungarian Academy of Sciences.}  J.K.'s research on gas-bearing circumstellar disks orbiting nearby young stars is supported by NASA Exoplanets program grant NNX16AB43G to RIT.  This paper makes use of the following ALMA data: ADS/JAO.ALMA\#2012.1.00195.S.  ALMA is a partnership of ESO (representing its member states), NSF (USA), and NINS (Japan), together with NRC (Canada) and NSC and ASIAA (Taiwan), in cooperation with the Republic of Chile.  The Joint ALMA Observatory is operated by ESO, AUI/NRAO, and NAOJ.  The NRAO is a facility of the NSF operated under cooperative agreement by Associated Universities, Inc.  We thank Wesleyan University for time on its high performance computing cluster supported by the NSF under grant number CNS-0619508.

\bibliographystyle{apj}
\bibliography{ms}

\end{document}